\documentclass[aps,prx,reprint,superscriptaddress,showpacs,preprintnumbers,amsmath,amssymb]{revtex4-1}
\usepackage[dvipdfmx]{graphicx}
\usepackage{dcolumn}   % needed for some tables
\usepackage{bm}        % for math
\usepackage{amssymb, amsmath, amsfonts}   % for math
\usepackage{lipsum}        % for math
\usepackage{color}
\DeclareGraphicsExtensions{{.pdf}}
\begin{document}
\newcommand{\noter}[1]{{\color{red}{#1}}}
\newcommand{\noteb}[1]{{\color{blue}{#1}}}
\newcommand{\field}{\left( \boldsymbol{r}\right)}
\newcommand{\paren}[1]{\left({#1}\right)}
\newcommand{\vect}[1]{\boldsymbol{#1}}
\newcommand{\uvect}[1]{\tilde{\boldsymbol{#1}}}
\newcommand{\vdot}[1]{\dot{\boldsymbol{#1}}}
\newcommand{\vder}{\boldsymbol{\nabla}}
\widetext
\title{
Unified View of Avalanche Criticality in Sheared Glasses
}
\author{Norihiro Oyama}
\email{oyamanorihiro@g.ecc.u-tokyo.ac.jp}
\affiliation{Graduate School of Arts and Sciences, The University of Tokyo, Tokyo 153-8902, Japan}
\affiliation{Mathematics for Advanced Materials-OIL, AIST, Sendai 980-8577, Japan}
\author{Hideyuki Mizuno}
\affiliation{Graduate School of Arts and Sciences, The University of Tokyo, Tokyo 153-8902, Japan}
\author{Atsushi Ikeda}
\affiliation{Graduate School of Arts and Sciences, The University of
Tokyo, Tokyo 153-8902, Japan}
\affiliation{Research Center for Complex Systems Biology, Universal Biology Institute, University of Tokyo, Komaba, Tokyo 153-8902, Japan}

\date{\today}
\begin{abstract}
Plastic events in sheared glasses are considered an example
of so-called avalanches, whose sizes obey a power-law probability
distribution with the avalanche critical exponent $\tau$.
Although mean-field theory predicts a universal value of
this exponent, $\tau_{\rm MF}=1.5$,
numerical simulations have reported different values depending on the literature.
Moreover, in the elastic regime, it has been noted that the
critical exponent can be different from that in the steady state,
and even criticality itself is a matter of debate.
Because these confusingly varying results were reported under different setups,
our knowledge of avalanche criticality in sheared glasses is greatly limited.
To gain a unified understanding, in this work,
we conduct a comprehensive numerical investigation of
avalanches in Lennard-Jones glasses under athermal quasistatic shear.
In particular, by excluding the ambiguity and arbitrariness that has crept into the
conventional measurement schemes,
{we achieve high-precision measurement and
demonstrate that the exponent $\tau$ in the steady state follows the mean-field prediction of
$\tau_{\rm MF}=1.5$.}
Our results also suggest that there are two qualitatively different avalanche
events.
This binariness leads to the non-universal behavior
of the avalanche size distribution and is likely to
be the cause of
the varying values of $\tau$ reported thus far.
{To investigate the dependence of criticality and universality on
applied shear,}
we further study the statistics of avalanches in the elastic
regime and
the ensemble of the first avalanche event in different samples, which
provide information about the unperturbed system.
We show that while the unperturbed system is indeed
off-critical, criticality gradually develops as shear is
applied.
The degree of criticality is encoded in the fractal dimension
of the avalanches, which starts from zero in the off-critical unperturbed
state and saturates in the steady state.
Moreover, the critical exponent obeys the mean-field
prediction $\tau_{\rm MF}$ universally, regardless of the amount of applied shear, once the system
{becomes critical}.
\end{abstract}
\maketitle
%
%%%%%%%%%%%%%%%%%%%%%%%%%%%%%%%%%%%%%%%%%%%%%%%%%%%%%%%%%%%%%%%%%%%%%%%%%%%%%%%%%%%%%%%%%%%%%%%%%%%%%%%%%%%%%%%%%%%%%%%%%%%%%%
% Introduction
%%%%%%%%%%%%%%%%%%%%%%%%%%%%%%%%%%%%%%%%%%%%%%%%%%%%%%%%%%%%%%%%%%%%%%%%%%%%%%%%%%%%%%%%%%%%%%%%%%%%%%%%%%%%%%%%%%%%%%%%%%%%%%
\section{Introduction}
%%% 1. General Introduction to Avalanche ``Universality''
It has been empirically accepted that various non-equilibrium systems
exhibit intermittent
fluctuations whose sizes obey a power-law distribution, $P(S)\sim
S^{-\tau}$, where $S$ is an appropriately defined size of intermittent
events and $\tau>0$ is
the critical exponent.
Such intermittent and scale-free fluctuations are called avalanches,
and the nature of the criticality of these fluctuations
{are expected to} form (sub-)classes of nonequilibrium universality~\cite{Sethna2001}.
Possible candidates for members of these classes cover a very wide range, including snow avalanches~\cite{Faillettaz2004} (as the name suggests),
Barkhausen noise~\cite{Sethna1993,Dahmen1996},
the depinning transition of elastic bodies moving in random media~\cite{Narayan1993,Fisher1998},
charge excitations in electron
glasses~\cite{Palassini2012},
microcrystal collapse under external forces~\cite{Dahmen2009},
earthquakes~\cite{Fisher1997,Dahmen1998},
the flickering of faraway stars~\cite{Sheikh2016}, the extinction of biological
species~\cite{Sole1996}, the firings of neuronal networks~\cite{DeArcangelis2006,Friedman2012}
and decision-making processes~\cite{Galam1997}.
Note that the theories of some of these examples
provide the same value of the critical exponent $\tau=1.5$,
at least at the mean-field level~\cite{Sethna1993,Fisher1998,Dahmen1998,Dahmen2009}

%%% 2. Avalanches in Amorphous Solids
Glasses under external fields such as shearing deformation or
compression, the target system of this article, have also been found
to exhibit intermittent
noise~\cite{Maloney2004,Aharonov2004,Tanguy2006,Bailey2007,Lerner2009,Tsamados2010,Heussinger2010b,Hatano2015,Oyama2019,AbedZadeh2019}.
In the case of sheared amorphous solids, the intermittency comes from plastic events.
The elementary process of plastic events is {believed to be so-called}
local shear transformation zones (STZs), which are triggered when the lowest eigenvalue of the dynamical matrix becomes zero~\cite{Maloney2004,Manning2011}.
STZs interact with each other via an elastic field, so the energy
released from an excited STZ can trigger further excitation of others~\cite{Maloney2006}.
Such a chain of STZs leads to scale-free avalanches.
{The theoretical treatment of avalanches} in sheared amorphous solids
has been achieved in a mean-field
manner\cite{Dahmen2011,Otsuki2014}, and
it has been shown that the critical exponent coincides with the universal
value $\tau_{\rm MF}=1.5$ reported
for other systems, such as Barkhausen noise~\cite{Sethna1993}, the
depinning transition of elastic bodies moving in random media~\cite{Fisher1998},
plastic events in deformed crystals~\cite{Dahmen2009}, and earthquakes~\cite{Dahmen1998}.
In particular, a general mean-field description of plasticity in
solids in ref.~\cite{Dahmen2009} can be applied to various solid systems,
such as crystals,
%Editor: On the line below, please ensure that the intended meaning has been maintained in this edit.
amorphous solids,
granular matter and earthquakes.

%%% 3. Experiments and numerical simulations
%% 3a. Experments
Many experimental and numerical studies have also been carried out.
Experimentally, {although there are still variations in
the precise value of $\tau$ depending on the literature~\cite{Sun2010,Bares2017},}
it has recently been reported that
the values of $\tau$ in various systems are universally close to
the mean-field value $\tau_{\rm MF}=1.5$~\cite{Antonaglia2014,Denisov2016,Tong2016,Murphy2019}.
In particular, since it is known that the precise value of $\tau$ is sensitive to the temporal resolution of the measurement~\cite{Leblanc2016} and
a study using a high resolution~\cite{Antonaglia2014} has reported a
value of $\tau$ consistent with $\tau_{\rm MF}$,
a consensus is being established---that the critical exponent in real systems
universally follows the mean-field prediction.

%% 3b. Simulations
For many situations,
numerical simulations can serve as powerful tools
that allow us to perform precisely controlled idealized
\emph{numerical experiments}.
In particular, simulations under an idealized condition have been performed to study avalanches in in sheared amorphous solids:
in most numerical works,
the limit of zero temperature and zero
strain rate, so-called athermal quasistatic (AQS) shear, is
employed~\cite{Kobayashi1980}.
Many studies have reported measurement results of
$\tau$ under AQS shear with various
setups that have included different frameworks---namely, atomistic simulations~~\cite{Salerno2012,Salerno2013,Zhang2017,Bares2017,Ozawa2018,Saitoh2019,Shang2020}
and elastoplastic
models~\cite{Talamali2011,Budrikis2013,Lin2014,Budrikis2017,Ferrero2019,Ferrero2019a}.
Some of these works further conducted finite-size scaling to validate the obtained value of $\tau$.
However,
the value of $\tau$ varies greatly from study to
study in the range of 1.0 to 1.36~\cite{Salerno2012,Salerno2013,Liu2016,Zhang2017,Bares2017,Ozawa2018,Saitoh2019,Shang2020,Talamali2011,Budrikis2013,Lin2014,Budrikis2017,Ferrero2019,Ferrero2019a}\footnote{
The maximum value becomes 1.5 if we also include systems under
oscillatory shear~\cite{Otsuki2014,Leishangthem2017}.}
(if we restrict the targets to only recent atomistic simulations, the range
becomes $\tau\in[1.0,1.3]$~\cite{Salerno2012,Salerno2013,Zhang2017,Shang2020}\footnote{Here, only works with finite-size
scalings are considered. Before these works, much smaller values were reported, such as in ref.~\cite{Maloney2004}. Additionally,
~\cite{Bares2017} and \cite{Saitoh2019} reported larger values. }).
Note that such variation is found even within the same
numerical framework.
In other words, even with the aid of idealized numerical experiments,
thus far, we have obtained only system-dependent values of critical exponents,
and no clues of universality or consistency with the mean-field
theory have been found.
This situation is at odds with that of experimental studies. 

%% 3c. Elastic regime
Furthermore, theoretically and numerically,
a new view has been proposed recently, making the situation even more confusing.
The new view states that avalanches in the elastic regime
show very different statistics than those in a steady state:
a mean-field replica theory specific to the elastic regime~\cite{Franz2017}
predicts that the critical exponent in this regime should be
$\tau_{\rm R}=1.0$ (if the system is \emph{above jamming}),
and a recent numerical work~\cite{Shang2020} has reported consistent
results in a binary Lennard-Jones (LJ) glass system.
The value of $\tau_{\rm R}=1.0$ is markedly smaller than the values
reported in the steady state, $\tau\in[1.15, 1.3]$~\cite{Salerno2012,Salerno2013,Zhang2017},
so the possibility of a change in the universality class
after the yielding transition takes place
has been suggested~\cite{Shang2020}.
However, even if we look at similar strain regimes,
different values ($\tau=1.1, 1.2$) that are consistent with the
results in the steady state have been reported in other
numerical works under AQS shear~\cite{Ozawa2018,Ruscher2019}.
Additionally, we highlight that experiments with high temporal resolution~\cite{Antonaglia2014} reporting
$\tau\approx \tau_{\rm MF}=1.5$ were conducted in the elastic regime.
Therefore, the value of the critical exponent in the elastic regime is still
under debate as well.
Moreover, ref.~\cite{Karmakar2010a} reported that, in the first place, the systems do not
exhibit criticality in the limit of $\gamma\to 0$, where $\gamma$ is
the accumulated applied shear strain.
Since all these seemingly conflicting results have been
reported under various numerical setups, we still lack a
firm understanding with a unified perspective.

%%% 4. Short summary of the presenting work
In this work, to resolve this puzzling situation concerning
avalanche criticality and universality presented above and to provide a unified view,
we investigate the statistics of avalanches in sheared
glasses comprehensively
by means of atomistic simulations of
binary LJ glasses under AQS simple shear.
First, by excluding the ambiguity and arbitrariness
that unexpectedly crept into the measurement of avalanche statistics in
previous works,
we show that the critical exponent $\tau$ in the steady state
coincides with the universal value obtained by the mean-field theory, $\tau_{\rm MF}=1.5$.
We stress that we obtain this value by using scaling relations,
not by a direct fitting of the data, which would require choosing the fitting range and thereby introducing unintentional arbitrariness.
Our results also suggest that the scaling function of the avalanche
size distribution has a peculiar bump and
thus is different from the one that we expected previously.
We find that there are two qualitatively different avalanche events,
which we call precursors and mainshocks.
Precursors and mainshocks follow different probability
distribution functions (PDFs), and
the peculiar bump of the scaling function
is found to be due only to the contribution from mainshocks;
these include system-spanning events and suffer from the finite-size effect.
Importantly, we also demonstrate that this bumpiness in the scaling function
explains the non-universal values of $\tau$ reported in
previous studies.

We then perform the same high-precision measurement in the
elastic regime to investigate whether we indeed observe
shear-dependent changes in criticality and universality.
In particular, we separately measure the statistics of both the ensembles of only
the initial avalanche events of different samples, which reflect the property of
the unperturbed system ($\gamma\to 0$), and the avalanches collected
in the elastic regime $0\le \gamma\le 0.02$~\cite{Shang2020}.
The former case does not exhibit any system size
dependence, in accordance with ref.~\cite{Karmakar2010a}.
Meanwhile, the latter case does show system size dependence, or
criticality, in agreement with ref.~\cite{Lin2015,Shang2020}.
This criticality in the elastic regime is clearly different from that in the steady state
and is characterized by a much smaller fractal dimension.
Nevertheless, consistent with the experimental results~\cite{Antonaglia2014,Denisov2016}, the value of $\tau\approx 1.471$ estimated by the scaling
relation is reasonably close to the steady-state value and $\tau_{\rm
MF}$.
{All these results provide a unified view of avalanche
criticality in sheared LJ glasses:
criticality develops as shear is exerted, and
the critical exponent $\tau$ remains the same universally
once the system becomes critical.
The development of criticality is reflected by the increasing
value of the fractal dimension, from zero in the off-critical
unperturbed system to a saturated value in the steady state.}

This article is organized as follows:
In Chapter~\ref{chap:method}, the numerical methods are summarized.
In particular, we introduce a new measurement scheme and
important scaling relations, including recapitulating those proposed in
ref.~\cite{Lin2014}.
The results for the steady state are presented in
Chapter~\ref{chap:results}.
In Chapter~\ref{chap:discussion},
the results of the elastic regime are presented, and
the unified view of avalanche criticality and universality throughout
the whole strain regime is provided.
Finally, concluding remarks are presented in
Chapter~\ref{chap:conclusion}.

%%%%%%%%%%%%%%%%%%%%%%%%%%%%%%%%%%%%%%%%%%%%%%%%%%%%%%%%%%%%%%%%%%%%%%%%%%%%%%%%%%%%%%%%%%%%%%%%%%%%%%%%%%%%%%%%%%%%%%%%%%%%%%
% Method
%%%%%%%%%%%%%%%%%%%%%%%%%%%%%%%%%%%%%%%%%%%%%%%%%%%%%%%%%%%%%%%%%%%%%%%%%%%%%%%%%%%%%%%%%%%%%%%%%%%%%%%%%%%%%%%%%%%%%%%%%%%%%%
\section{Methods}\label{chap:method}
In this work, we conduct simulations of two-dimensional ($d=2$) sheared binary
LJ glasses and investigate the avalanche statistics in detail.
Specifically, we aim to exclude ambiguities from the definition and
measurement of avalanche sizes.
In this chapter, we first explain the numerical setup of our binary LJ
glass system under AQS shear in Sec.~\ref{sec:system}.
In Sec.~\ref{sec:avalanche_definition},
we propose a brand-new measurement scheme for avalanches.
In the subsequent section, Sec.~\ref{sec:resolution}, we discuss the
importance of system size-dependent tuning of the numerical strain
interval $\Delta \gamma$, which has not been taken seriously thus far.
Finally, the scaling relations proposed in ref.~\cite{Lin2014} are
summarized in Sec.~\ref{sec:scaling} in a way that is compatible with
our setup.
%%%
\subsection{Target system}\label{sec:system}
%%%
%%%{Additive Smoothed Lennard-Jones Potential}
For the inter-particle potential, we employ the smoothed
LJ potential~\cite{Salerno2013}, defined as
\begin{align}
  \phi_{\rm L}(r_{ij}) &= 4\epsilon_{ij}\left[ \left( \cfrac{d_{ij}}{r_{ij}}\right)^{12}-\left( \cfrac{d_{ij}}{r_{ij}}\right)^{6}\right] + \epsilon_{\rm C}\quad (r_{ij}<I_{ij})\\
  \phi_{\rm R}(r_{ij}) &= \cfrac{C_3}{3}(r_{ij}-r_{ij}^{\rm
    C})^3+\cfrac{C_4}{4}(r_{ij}-r_{ij}^{\rm C})^4\quad (I_{ij}\le
  r_{ij}<r^{\rm C}_{ij})
\end{align}
where $r_{ij}$ is the inter-particle distance between particles $i$
and $j$, $d_{ij}$ determines the interaction range, and
$\epsilon_{\rm C}$ is the
potential offset,
{which guarantees that $\phi_{\rm L}$ and $\phi_{\rm
R}$ (and their first and second derivatives) match
at the inner cutoff $I_{ij}\equiv 1.2d_{ij}$.
The coefficients $C_3$ and $C_4$ are chosen so that
$\phi_{\rm R}$ and its first and second derivatives continuously go to zero at the outer cutoff
$r_{ij}^{\rm C} \equiv 1.3d_{ij}$.}
%%%
%% Binary Mixture
To avoid crystallization, the system is composed of two different
sizes of particles, species S and L, at a ratio of $50:50$.
The potential is totally additive, and the interaction ranges are
$d_{\rm SS}=5/6$, $d_{\rm
SL}=1.0$ and $d_{\rm LL}=7/6$, respectively.
The energy unit $\epsilon_{ij}=\epsilon = 1.0$ is constant for all
combinations of particle species.
Below, all physical variables are non-dimensionalized by the length unit
$d_{SL}$ and the energy unit $\epsilon$.
The number density of the system is fixed at
$\rho=N/L^2\approx 1.09$.
All samples are generated by minimizing the potential energy of a completely random
initial configuration, which corresponds to an infinite temperature.

%%%{Athermal Quasistatic (AQS) Shear}
The system is driven out of equilibrium by external simple shear.
The simple shear is imposed on the whole system in a quasistatic
way without any thermal noise.
This protocol is called AQS shear and is
achieved by the repetition of very tiny affine shearing deformations of the
strain increment $\Delta \gamma$, followed by energy minimization
under the Lees-Edwards boundary condition~\cite{Allen1987}.
The energy is considered to be minimized when the maximum
magnitude of the forces applied to the particles $f_{\rm max}$ meets the condition $f_{\rm
max}<10^{-9}$.
We use
the FIRE algorithm
for energy minimization~\cite{Bitzek2006}.
Although several works have reported that the introduction of inertia
during energy minimization can affect the avalanche statistics,
such an effect seems to be absent in our results (see Appendix~\ref{ap:bumps}).
Even under these conditions, we still have one free parameter---namely,
the strain resolution per numerical step $\Delta \gamma$.
In this work, we tune this parameter depending on the system size
$N$, and this tuning plays a fundamental role in measuring the avalanche
exponent $\tau$.
The determination of $\Delta \gamma$ will be discussed in
Sec.~\ref{sec:resolution}.

%%%
\subsection{Rewinding method and the definition of avalanches}\label{sec:avalanche_definition}
\begin{figure}
  \includegraphics[width=0.8\linewidth]{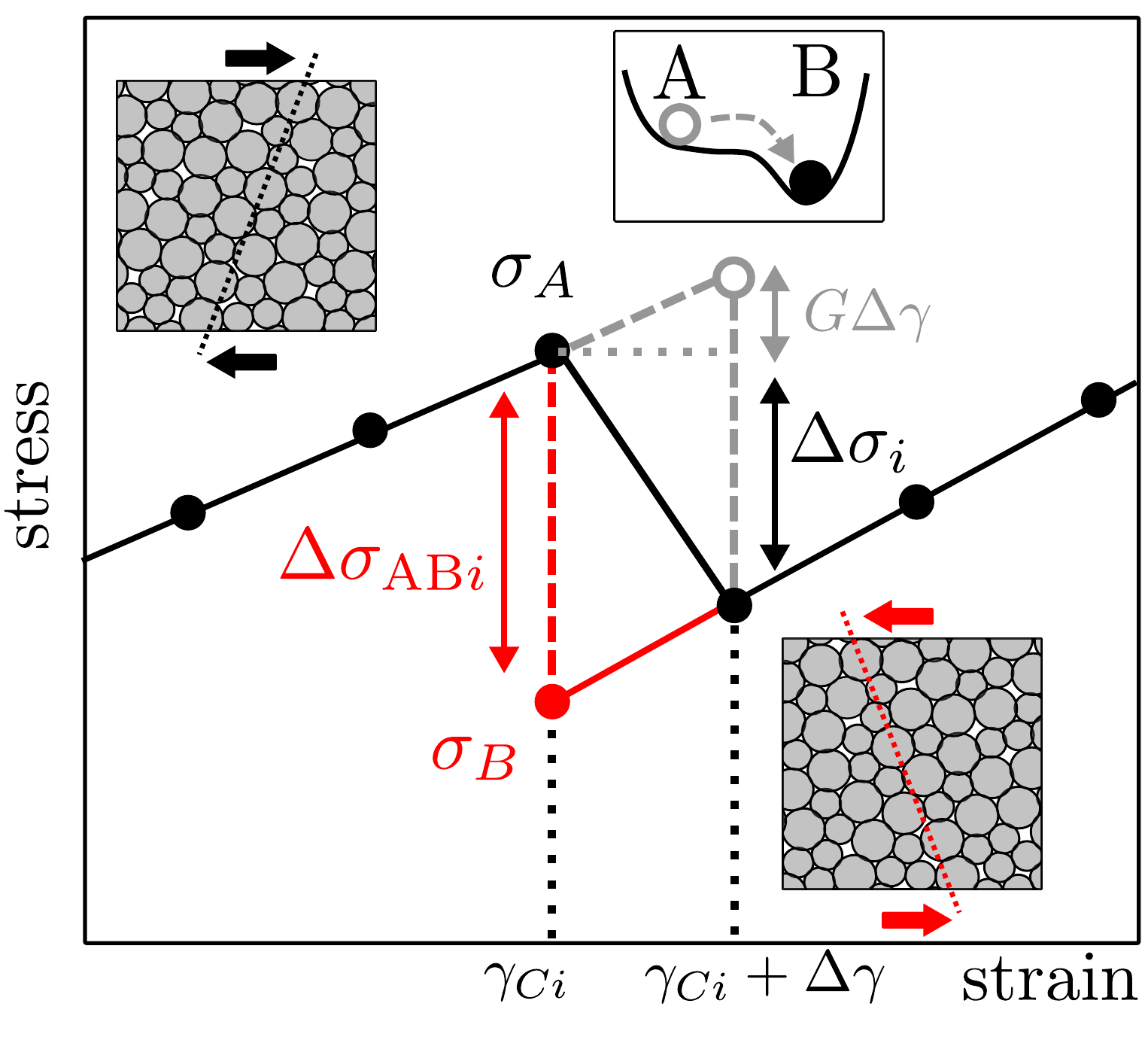}
\caption{
Schematic picture of the rewinding method.
The black lines and points represent the results of a normal AQS
simulation.
The gray lines and points depict the conventional definition of the
avalanche size with a linear correction.
The red lines and points represent our definition of the avalanche size
using the rewinding method.
    \label{fig:rewind}}
\end{figure}

To
evaluate the size of avalanches that are purely due to plastic events,
stress drops (or potential energy drops)
should be measured under the same boundary conditions.
For this reason, in previous studies~\cite{Bailey2007,Zhang2017,Ozawa2018,Shang2020},
the size of the $i$th avalanche $S_i$ is defined as the sum of the stress drop and
linear correction, as
\begin{align}
  S_i\equiv L^d (\Delta \sigma_i + G\Delta \gamma),\label{eq:LCorre}
\end{align}
where $\Delta \sigma_i\equiv
\sigma(\gamma_{Ci})-\sigma(\gamma_{Ci}+\Delta\gamma)$ is the stress
drop during the $i$th avalanche, $\gamma_{Ci}$ is the critical strain
at which the $i$th avalanche takes place, and $G$
is the shear modulus (Fig.~\ref{fig:rewind}).
However, as discussed in ref.~\cite{Maloney2004},
the value of the shear modulus $G$ fluctuates strongly when an
external shear is applied.
In particular, $G$ becomes infinite in the negative direction at the onset of
an avalanche where the lowest eigenvalue of the dynamical
matrix becomes zero.
It is even possible that a single stress drop event can take several
numerical strain steps when the strain resolution $\Delta \gamma$ is
very fine~\cite{Maloney2004}.
Therefore, it is quite nontrivial to determine which kind of definition of the
modulus should be used for the linear correction in
Eq.~\ref{eq:LCorre} and how different definitions affect the results.

%------------------------------------------------------------------------------------------
%%% Figure: Schematic picture of Rewind Method
%------------------------------------------------------------------------------------------
%------------------------------------------------------------------------------------------

To rule out such an ambiguity in the definition of avalanche sizes,
we developed a new measurement scheme:
when a stress drop event is detected, we
reverse the direction of shear and rewind the strain by one strain step $\Delta \gamma$
(see Fig.~\ref{fig:rewind}).
We call this scheme \emph{the rewinding method}.
From the perspective of the potential energy landscape picture,
a plastic event can be viewed as a transition from one metabasin to another; call them states A (original) and B (new).
{The rewinding method enables us to directly compare the variables of these
two states A and B at exactly the same boundary condition $\gamma=\gamma_{Ci}$.}
Thus, we can define the $i$th avalanche size $S_i$ simply by the difference
between the stresses of the two states without any ambiguity, as
\begin{align}
  S_i = L^d\Delta\sigma_{{AB}i},\label{eq:our_def}
\end{align}
where $\Delta\sigma_{{AB}i}\equiv \sigma_A(\gamma_{Ci})-\sigma_B(\gamma_{Ci})$ and $\sigma_s(\gamma)$ denotes the stress of state $s\in {A,B}$ at strain
$\gamma$.
Hereafter, all our analyses are based on this definition,
Eq.~\ref{eq:our_def}.
We emphasize that we do not introduce any lower cutoff size for avalanche
detection~\cite{Ozawa2018,Shang2020}, and we utilize all stress drop events in this work.

%%%
\subsection{Strain resolution}\label{sec:resolution}
%------------------------------------------------------------------------------------------
%%% Table: Values of strain resolution
%------------------------------------------------------------------------------------------
\tabcolsep = 5pt
\begingroup
\renewcommand{\arraystretch}{1.5}
\begin{table}[tb]
  \caption{Values of the strain resolution}
  \centering\label{table:resolution}
  \begin{tabular}{c|ccccc}
    \hline
    $N$ & 512 & 2048  & 8192 & 32768  & 131072*\\
    \hline 
    $\Delta \gamma$ &$5\times 10^{-6}$  &$5\times 10^{-6}$  &$5\times
    10^{-6}$  &$1\times 10^{-6}$ &$5\times 10^{-7}$ \\
    \hline
    \multicolumn{6}{c}{*this size is considered for only the first event ensemble}
  \end{tabular}
\end{table}
\endgroup

%%%%%%%%%%%%%%%%%%%%%%%%%%%%%%%%%%%%%%%%
Since the average of the strain intervals between
avalanches, $\delta \gamma_i\equiv \gamma_{Ci+1}-\gamma_{Ci}$, is known to decrease with increasing system size $N$ as
$\langle\delta\gamma\rangle\sim N^{-\chi}$ with a positive exponent $\chi$~\cite{Karmakar2010a,Karmakar2010,Lin2014},
larger systems require finer strain resolutions to detect small 
avalanches properly.
In other words, if we do not care about the
strain resolution $\Delta \gamma$,
the statistics of small avalanches in large systems
can be obscured.
However, thus far, in most cases, $\Delta \gamma$ has been
more or less fixed to a single value regardless of the system size.
Alternatively, a smallest size cutoff for avalanche detection has sometimes been
introduced~\cite{Ozawa2018,Shang2020}.
Such treatments would be justified if the
scale-free power-law behavior appeared in the large size regime of the PDF
of avalanche sizes
close to the cutoff size $S_c$.
As discussed later, however, our results show that this is not the case and indicate
the importance of tuning the strain resolution $\Delta \gamma$
depending on the system size.
The precise values of the strain resolution $\Delta\gamma(N)$ that we used
for the different system sizes
are summarized in Table~\ref{table:resolution}.
See Appendix~\ref{ap:validation} for how we determined these values and
how the results are affected if we do not tune $\Delta \gamma$ properly.

%%%
\subsection{Scaling laws}\label{sec:scaling}

%------------------------------------------------------------------------------------------
%%% Table: Lists of related critical exponents
%------------------------------------------------------------------------------------------
\tabcolsep = 15pt
\begingroup
\renewcommand{\arraystretch}{1.5}
\begin{table*}[htb]
  \caption{List of the critical exponents}
  \centering\label{table:exponents}
  \begin{tabular}{cccc}
    \hline
    Exponent & Definition  & Estimation in this work & Steady-state value\\
    \hline 
    % chi
        {$\chi$} & $\left < \Delta \gamma \right > \sim N^{-\chi}$  &
        Direct fitting
        & 0.738\\
        % d_f
        {$d_f$} & $S_c\sim L^{d_f}$ & Direct fitting & 1.034\\
        % theta
        {$\theta$} &$P(x) \sim x^{\theta}$  & $\theta=(1-\chi)/\chi$ & 0.355\\
        % tau
        $\tau$ & $P(S)\sim S^{-\tau}$ & $\tau = 2-(1-\chi){d}/{d_f}$ & 1.493\\
        % alpha
        {$\alpha$} & $\langle S \rangle \sim
        N^{\alpha}$ & {$\alpha=(2-\tau)d_f/d=1-\chi$} & 0.262\\
        % beta
        $\beta$ & $R(X,L)=L^\beta g(X/L^{d_f})$  &
        $\beta=d\chi-d_f\tau=d-2d_f$ & -0.068\\
        % gamma
        %$\gamma$ & $R(X,L)=L^\gamma g(X)$ & $\gamma=d\chi$ &1.446\\
        %& & from $\gamma=\beta+\alpha\tau$\\
        \hline
  \end{tabular}
\end{table*}
\endgroup

%%%%%%%%%%%%%%%%%%%%%%%%%%%%%%%%%%%%%%%%

In this section, we summarize the scaling relation, which
reduces the number of independent critical exponents through physical
constraints.
In particular, by following the original discussion in
ref.~\cite{Lin2014},
we show that these relations are closed by
only two exponents.
We list all six related critical exponents
in Table~\ref{table:exponents}, and to ensure that the article is self-contained, we
recapitulate the derivations of all the relations.
The whole discussion here relies largely on the concept of
marginal stability and one of its consequences, the pseudogap in the PDF
of the local stability.
We briefly summarize
these concepts in Appendix~\ref{ap:pseudogap}.

{Although 
the pseudogap exponent $\theta$ plays an important role
in the current context,
the PDF of the local stability $P(x)$, through which we can obtain
$\theta$, cannot be measured directly in
particle-based simulations; $x$ stands for local stability
(see Appendix~\ref{ap:pseudogap} for the precise definition of the
pseudogap exponent $\theta$ and the local stability $x$).
Instead, $\theta$ is measured indirectly as follows:
Since a plastic event is excited when the applied shear equals the
minimum value of local stability $x_{\rm min}$, we trivially
obtain the following relation only if the
local stabilities $x_i$ are independent:}
\begin{align}
  \int_0^{\langle x_{\rm min}\rangle}P(x)&\sim \frac{1}{N},\\
  \Leftrightarrow \langle{x_{\rm min}}\rangle &\sim N^{-1/(\theta+1)}.
\end{align}
Since $x_{\rm min}$ corresponds to the strain
intervals $\delta \gamma_i$ in the current setup,
we obtain the first scaling relation, which allows us to estimate the
pseudogap exponent $\theta$ from the exponent $\chi$:
\begin{align}
  N^{-1/(\theta+1)}&\sim \langle\delta \gamma\rangle\sim N^{-\chi},\\
  \Leftrightarrow \theta &= \frac{1-\chi}{\chi}.
\end{align}

We now introduce
two different PDFs of avalanche sizes.
The first, $P(S)$, is the standard normalized PDF per unit avalanche
size and is simply given as
\begin{align}
  \int P(S)dS =1.
\end{align}
The other, $R(S)$, is the PDF per unit avalanche size and unit
strain.
If we define the average number of avalanche events per unit strain
$M(L)$ as
\begin{align}
  M(L) &=\int_0^{\infty} R(S,L) dS,
\end{align}
then these three functions can be related to each other as
\begin{align}
  P(S, L) &= R(S, L)/M(L),\label{eq:R_definition}
\end{align}
where $L$ is the linear dimension of the system. We now explicitly write the system size dependence of the
PDFs.
In the field of avalanches in sheared glasses~\cite{Salerno2012,Salerno2013,Zhang2017,Shang2020},
the PDF per unit strain $R(S)$ is usually preferred to the standard
PDF $P(S)$.

We now assume the criticality of avalanches and that the
distribution has a system size-dependent cutoff size $S_c\sim
L^{d_f}$,
where $d_f$ is the fractal dimension.
Then, by introducing a
scaling function $f(S/S_c)$ as $P(S)=S^{-\tau}f(S/S_c)$,
we obtain a scaling relation for $P(S)$:
\begin{align}
  P(S)&\sim L^{-d_f\tau}(S/L^{d_f})^{-\tau}f(S/L^{d_f}),\\
  &\sim L^{-d_f\tau}g(S/L^{d_f})\label{eq:P_scaling},
\end{align}
where we introduce another function, $g(S/S_c)\equiv (S/S_c)^\tau f(S/S_c)$.
If we substitute $M(L)\sim 1/\langle \delta \gamma (L)\rangle\sim N^\chi$
and Eq.~\ref{eq:P_scaling} into Eq.~\ref{eq:R_definition}, we obtain
\begin{align}
  R(S) &\sim L^{d\chi-d_f\tau} g(S/L^{d_f}),\label{eq:R_scaling}
\end{align}
Thus, comparing Eq.~\ref{eq:R_scaling} with the definition of the
exponent $\beta$ shown in Table~\ref{table:exponents}, we obtain the
following relation:
\begin{align}
  \beta=d\chi-d_f\tau.\label{eq:beta_scaling1}
\end{align}

Another relation among $\tau$, $\chi$ and $d_f$ can be derived
from the stationary condition of stress in the steady state.
For $1<\tau<2$ (which is the case for avalanches in sheared
glasses), the average avalanche size can be derived from
$P(S)\sim S^{-\tau}$ as
\begin{align}
  \langle S\rangle\sim S_c^{2-\tau}\sim
  L^{d_f(2-\tau)}.\label{eq:ave_S1}
\end{align}
In the steady state, this value of $\langle S\rangle\equiv L^d \langle \Delta
\sigma\rangle$ must be consistent with the average increase in the stress
between avalanches.
Assuming that the average shear modulus $\bar{G}$ does not depend on
the system size statistically, this condition leads to
\begin{align}
  \langle S \rangle =L^d \bar{G}\langle \delta \gamma\rangle\sim L^{d\theta/(\theta+1)}.\label{eq:ave_S2}
\end{align}
From Eqs.~\ref{eq:ave_S1} and \ref{eq:ave_S2}, we obtain the relation
\begin{align}
  \tau&=2-\frac{\theta}{\theta+1}\cdot\frac{d}{d_f}=2-(1-\chi)\cdot\frac{d}{d_f}, \label{eq:Lin_scaling}
\end{align}
which plays a central role.

Eqs.~\ref{eq:beta_scaling1} and \ref{eq:Lin_scaling}
allow us to write $\beta$ in a simpler way:
\begin{align}
  \beta = d-2d_f.\label{eq:beta_scaling2}
\end{align}
Comparing Eq.~\ref{eq:ave_S1} and the definition of the exponent
$\alpha$ shown in Table~\ref{table:exponents}, and substituting
Eq.~\ref{eq:Lin_scaling}, we can express $\alpha$ as
\begin{align}
  \alpha=(2-\tau)d_f/d = 1-\chi.\label{eq:alpha_relation}
\end{align}

All these scaling relations ultimately reduce the number of independent exponents to two.
Therefore, we must select two independent exponents and describe
others using them.
We employ $d_f$ and $\chi$ in this work.
We stress that, while we must choose the fitting range to obtain
$\tau$ by direct fitting to the avalanche size distribution, the
relations $S_c\sim L^{d_f}$ and
$\langle \delta \gamma\rangle\sim N^{-\chi}$ are valid for the whole
data range, and $d_f$ and $\chi$ can be measured without any
arbitrariness in the choice of the fitting range.

%%%%%%%%%%%%%%%%%%%%%%%%%%%%%%%%%%%%%%%%%%%%%%%%%%%%%%%%%%%%%%%%%%%%%%%%%%%%%%%%%%%%%%%%%%%%%%%%%%%%%%%%%%%%%%%%%%%%%%%%%%%%%%
% Results
%%%%%%%%%%%%%%%%%%%%%%%%%%%%%%%%%%%%%%%%%%%%%%%%%%%%%%%%%%%%%%%%%%%%%%%%%%%%%%%%%%%%%%%%%%%%%%%%%%%%%%%%%%%%%%%%%%%%%%%%%%%%%%
\section{Statistics of avalanches in the steady state}\label{chap:results}
In this chapter,
we present the avalanche statistics in the steady state ($\gamma>0.25$).
For all system sizes, we collected more than 5000 events and
calculated the statistical information from them.
%%%
\subsection{Independent exponents}
\begin{figure}
  \includegraphics[width=\linewidth]{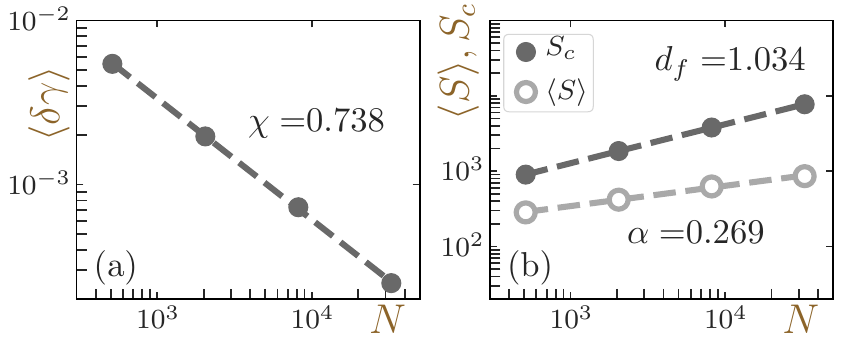}
\caption{
The system size dependence of (a) the average strain interval
between successive avalanches $\langle \delta\gamma\rangle$ and (b) the
mean and cutoff avalanche sizes, $\langle S \rangle$ and $S_c$,
in the steady state.
The markers represent the numerical results, and the dashed lines are power-law fittings.
The values of the exponents estimated from the fittings are also shown.
    \label{fig:steady_definite}}
\end{figure}

We start with the measurement of two independent exponents
$d_f$ and $\chi$, which determine all other exponents through the
scaling relations introduced in Sec.~\ref{sec:scaling}.
By definition, these two exponents can be measured from the system size
dependence of the average strain interval between avalanches $\langle
\delta \gamma\rangle$ and the cutoff avalanche size $S_c\equiv\langle S^2\rangle
/ \langle S\rangle$~\cite{Shang2020}.
As shown in Fig.~\ref{fig:steady_definite}, both $\langle
\delta \gamma\rangle$ and  $S_c$ are power-law functions of the system
size $N$, as expected.
We note that, as discussed in Appendix~\ref{ap:validation},
if we do not tune $\Delta\gamma$ carefully, $\langle \delta \gamma\rangle$
will not be a power-law function.
The obtained exponents and the estimation of the other exponents
are summarized in Table~\ref{table:exponents}.
In Fig.~\ref{fig:steady_definite}(b), we also show the results for $\langle
S\rangle$, which yields the exponent $\alpha$.
We note that the direct measurement result $\alpha=0.269$ shows very
close agreement with the estimation by Eq.~\ref{eq:alpha_relation}, {$\alpha\approx 0.262$}.
This supports the accuracy of the calculations and the scaling relation.

From Eq.~\ref{eq:Lin_scaling}
and the values of $d_f$ and $\chi$, the avalanche exponent is
estimated as $\tau\approx 1.493$.
We would like to emphasize that this value is very close to the mean-field prediction of $\tau_{\rm MF}=1.5$~\cite{Dahmen2009}.
To strictly confirm that our result for $\tau$ is the intrinsic
critical exponent of the system,
we conduct further validation in the next two sections.

%------------------------------------------------------------------------------------------
%%% Figure: Definite exponents in the steady state
%------------------------------------------------------------------------------------------
%------------------------------------------------------------------------------------------

%%%
\subsection{Avalanche size distribution}
\begin{figure}
  \includegraphics[width=0.7\linewidth]{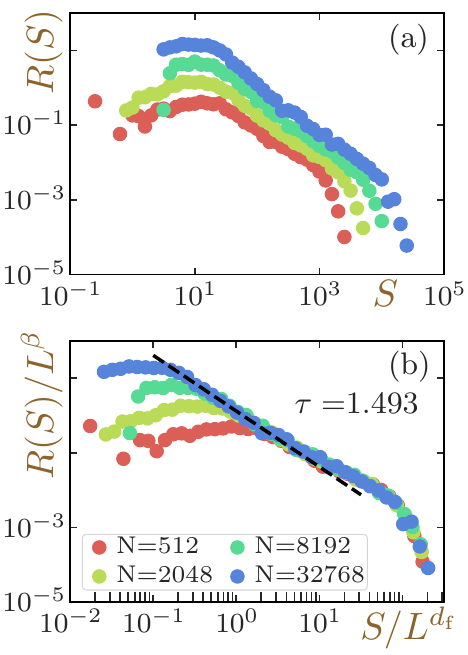}
\caption{
Unit strain probability distribution of avalanche sizes $R(S)$ in the
steady state
(a) without scaling; (b) with finite-size scaling.
The scaling exponents $d_f$ and $\beta$ are drawn from Table~\ref{table:exponents}.
For both panels, different colors are used for different system sizes, as shown in the
legend in (b).
The dashed line in (b) shows the power law
behavior predicted by the scaling law, Eq.~\ref{eq:Lin_scaling}, not a direct fitting result.
    \label{fig:steady_ava}}
\end{figure}

We now turn our attention to the avalanche size distribution.
Fig.~\ref{fig:steady_ava}(a) shows the results of the PDFs of the
avalanche size per unit strain $R(S)$ for different system sizes.
In Fig.~\ref{fig:steady_ava}(b), the same data are shown with a
finite-size scaling with the exponents $d_f$ and $\beta$.
{Here, we demonstrate that 
the exponent $\beta$ that is estimated solely from $\chi$ and $d_f$
without any further fitting leads to a collapse of the results of
different system sizes.}
This success of the collapse again supports the validity of our numerical calculation
and the scaling laws.

We intuitively expect that the power-law behavior should appear in the
large size regime near the cutoff size $S_c$, and in fact, previous
works have estimated the value of $\tau$ by a direct fitting of the data in that
regime~\cite{Salerno2012,Salerno2013,Zhang2017,Ozawa2018,Shang2020}.
However,
in Fig.~\ref{fig:steady_ava}(b), we see the
power-law regime, $S^{-\tau}$ with $\tau\approx 1.493$,
which the scaling relations suggest is located instead in the small size
regime.
We would like to stress that this regime actually grows broader as the system
size increases.
{This result suggests that the PDFs have peculiar bumps in the large
size regime; we now will reveal the origin of these bumps.}

{Before proceeding to the next section, we highlight that our value of $\tau\approx 1.493$ is much larger than the
values reported in previous works with atomistic simulations~\cite{Salerno2012,Salerno2013,Zhang2017}.}
In Appendix~\ref{ap:comparison1}, we show that if we estimate
the value of $\tau$ in the same way as in previous works---namely, by a direct fitting of our
data in the bumpy regime---we obtain $\tau\sim 1.19$.
This value lies in the middle of those
reported for the steady state in
previous works, $\tau\in[1.15,1.3]$.
{Based on this consistency,
we believe that the values of $\tau$ in previous
works varied widely only because the non-universal crossover regime
was analyzed.}

%------------------------------------------------------------------------------------------
%%% Figure: avalanche distribution and its finite size scaling
%------------------------------------------------------------------------------------------
%------------------------------------------------------------------------------------------

%%%
\subsection{Origin of the bump in the large size regime}
\begin{figure}
  \includegraphics[width=0.8\linewidth]{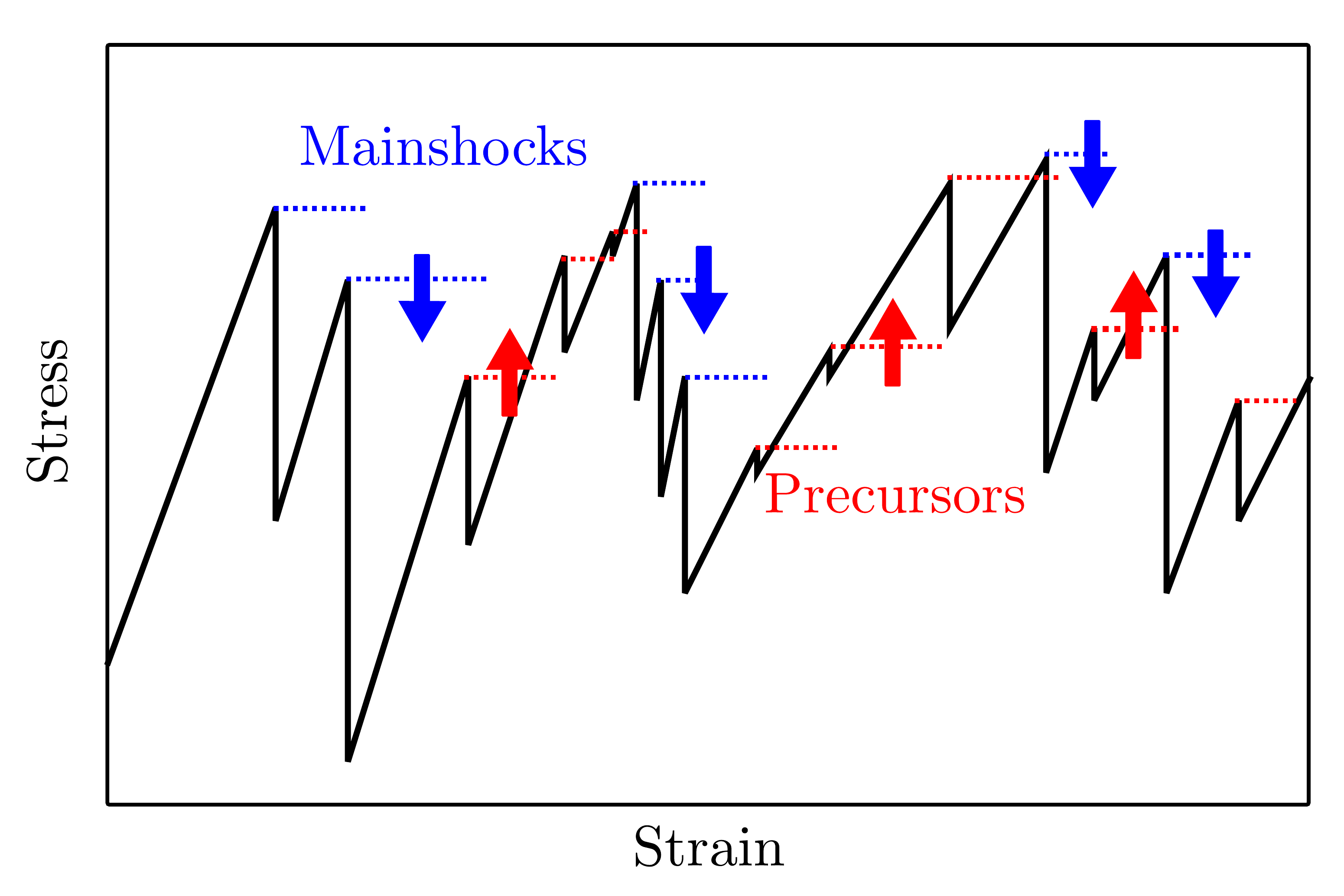}
\caption{
Schematic of the definition of precursors and mainshocks.
They are defined based on the relation between the stress
just before the event of interest and the next one.
The black lines depict the schematic stress-strain curve.
The dotted lines compare the stresses immediately before successive events.
The blue lines do not cross the stress-strain curve or mean downhill
mainshocks, while the red lines
cross them and indicate uphill precursors.
    \label{fig:pre_main_ske}}
\end{figure}

\begin{figure}
  \includegraphics[width=0.7\linewidth]{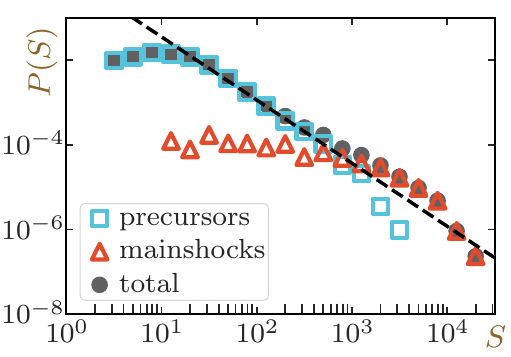}
\caption{
Decomposition of the PDF of the avalanche sizes in the steady state
into contributions from precursors and
mainshocks.
The PDFs of only precursors or mainshocks are normalized to the total
number of events.
The results for the system with $N=32768$ are shown.
The squares represent precursors, the triangles represent mainshocks, and the
circles are the results of the total distribution, as shown in the
legend.
The dashed line is a power-law relation with the exponent shown in
Fig.~\ref{fig:steady_ava}.
    \label{fig:steady_pre_main}}
\end{figure}

The bumpy nature of the PDF suggests that
the distribution is composed of two qualitatively different
contributions: scale-free power-law events and system size-dependent
percolated events.
We discovered qualitatively different groups of avalanche events that
prove this hypothesis.
As sketched in Fig.~\ref{fig:pre_main_ske},
the evolution of the macroscopic stress under the AQS shear exhibits qualitatively
different stress drop events---namely, uphill events and downhill events.
Whether an event of interest is uphill or downhill is judged
according to the relation between the stress immediately before the event
and that of the next one, $\sigma(\gamma_{Ci})$ and
$\sigma(\gamma_{Ci+1})$.
If $\sigma(\gamma_{Ci})$ is smaller (larger) than $\sigma(\gamma_{Ci+1})$,
the event of interest is considered to be an uphill (downhill) event.
We call uphill events precursors and downhill events mainshocks
hereafter.
Note that here, we define precursors and mainshocks without introducing any parameters.
In Fig.~\ref{fig:steady_pre_main}, we show that the PDF of the avalanche
sizes can be decomposed into contributions from only
precursors and mainshocks.
{Furthermore, the results show that the bump is purely
composed of mainshocks and that
the PDF of precursors obey a standard power-law behavior with a specific cutoff size.
This also suggests that the PDF of 
mainshocks includes an excess of large system-spanning events due to
the finite-size effects in addition to 
unbounded scale-free events that obey the same power-law PDF
as the precursors.}

%------------------------------------------------------------------------------------------
%%% Figure: Definition of Precursors/Mainshocks
%------------------------------------------------------------------------------------------
%------------------------------------------------------------------------------------------

%------------------------------------------------------------------------------------------
%%% Figure: Pre/Main separation of Avlanche/MSD distribution
%------------------------------------------------------------------------------------------
%------------------------------------------------------------------------------------------

\begin{figure}
  \includegraphics[width=\linewidth]{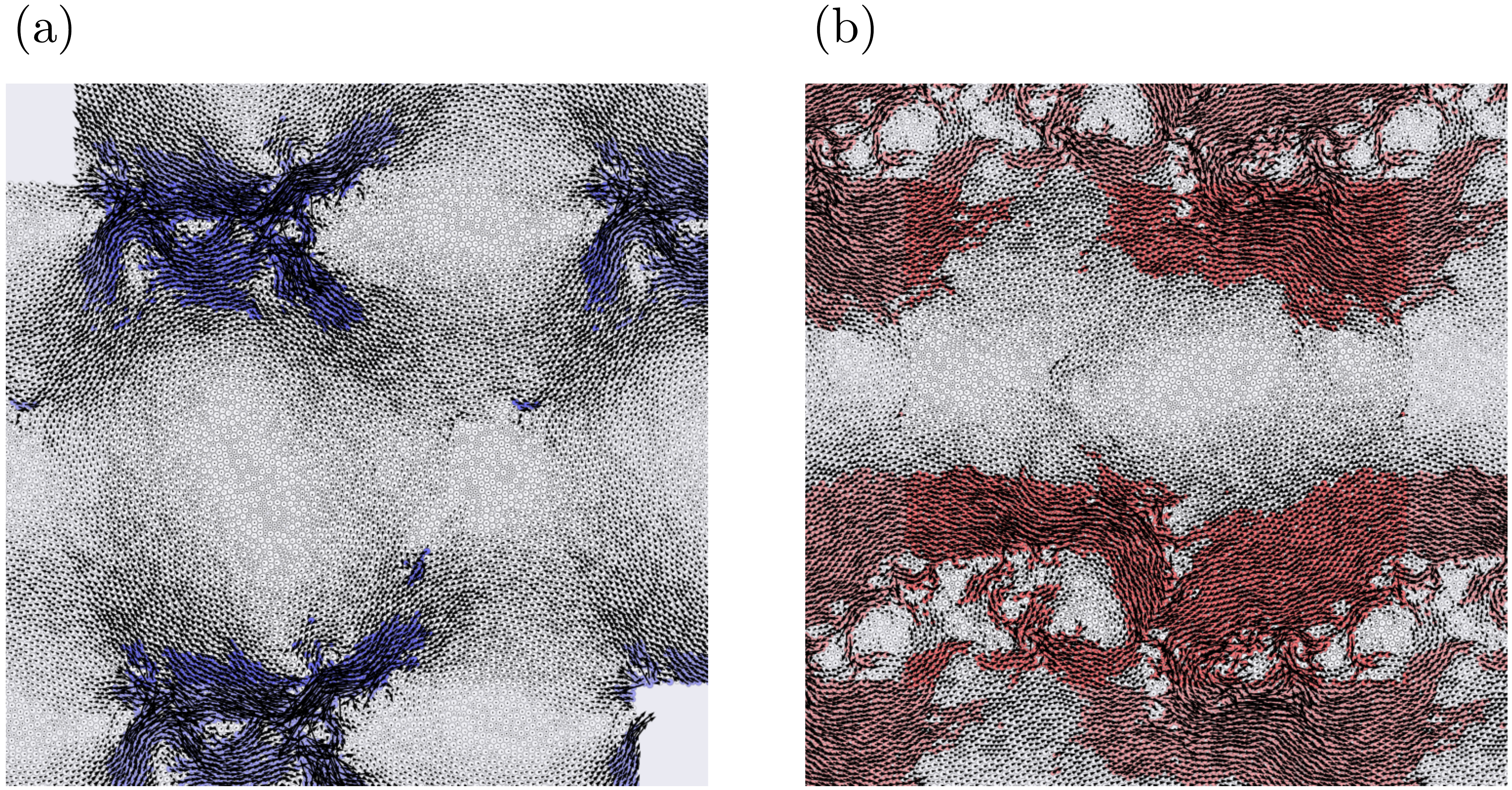}
\caption{
Visualization of the displacement field during an avalanche event in
a system with $N=8192$:
(a) precursor; (b) mainshock.
Events with the largest avalanche sizes for the same event types are shown.
The arrows represent the displacement vectors of particles and have been normalized
properly for ease of viewing.
The colored particles are mobile particles (see the main text for the definition).
The copied cells due to the periodic boundary conditions are shown around
the original cell with slightly lighter colors.
    \label{fig:visualization}}
\end{figure}

{The direct visualization of the displacement field during
a precursor and a mainshock provide
more insight into the difference between the two event types
(see Fig.~\ref{fig:visualization}).
In particular, the events with the largest avalanche sizes for the
same event types are shown.
Here, we highlight only mobile particles that 
are defined according to the participation ratio $e\equiv (\sum_i d_i^2)^2/(N\sum_id_i^4)$,
where $d_i=|\boldsymbol{d}_i|$ is the magnitude of the displacement
vector of particle $i$.
The participation ratio $e$ provides the fraction of particles that 
are mobile: if all displacement vectors have the same magnitude, 
$e=1$ holds, and if only one
vector has a nonzero value, $e=1/N$ holds.
We define particles that have the $e$ largest magnitudes of
displacement vectors as mobile particles.}
As shown in Fig.~\ref{fig:visualization},
even in the largest event, the mobile particles of a precursor exhibit a
localized structure,
while the mainshock counterpart is system-spanning due to
the finite-size effect.
Therefore, it is reasonable that precursors are mainly responsible for the
intrinsic scale-free power-law regime.
We stress that the event in Fig.~\ref{fig:visualization}(a) is
a chain of multiple STZs.

The qualitative features of mainshocks that have been presented so far are
reminiscent of the so-called runaway events observed in the mean-field
model of amorphous solids~\cite{Dahmen1998,Dahmen2009}.
In the mean-field model, such runaway events are expected
only when \emph{the weakening parameter} is positive---in other words,
when the
system shows brittle responses to an external shear, like metallic
glasses.
This may seem reasonable, since LJ glasses are sometimes used
as a model system of metallic glasses~\cite{Kob1995}.
However, our mainshocks exhibit qualitatively different scaling
behavior from runaway events.
In ref.~\cite{Barbara1991}, it was reported that runaway events cannot
be collapsed by the same scaling exponents as those for the power-law regime.
In our case, on the other hand, entire PDFs can be collapsed by a single
combination of scaling exponents, as presented in Fig.~\ref{fig:steady_ava}.
In this sense, our mainshocks are qualitatively different from runaway
events.
It is also important to mention that several studies have
reported that similar bumps in the
PDFs of avalanche sizes can be induced by the inertial effect~\cite{Salerno2012,Salerno2013,Karimi2017}.
We emphasize that they seem to be different in nature from our mainshocks.
This issue is discussed in detail in Appendix \ref{ap:bumps}.
We would like to note that we are aware of qualitatively similar bumps
in PDFs shown in previous works--- in both experimental~\cite{Sun2010} and numerical works~\cite{Lin2014a,Lerner2014a,Zhang2017}.
We emphasize that some of them are measured in completely inertialess conditions.

%------------------------------------------------------------------------------------------
%%% Figure: Visualization
%------------------------------------------------------------------------------------------
%\begin{figure*}
%\end{figure*}
%------------------------------------------------------------------------------------------

%%%%%%%%%%%%%%%%%%%%%%%%%%%%%%%%%%%%%%%%%%%%%%%%%%%%%%%%%%%%%%%%%%%%%%%%%%%%%%%%%%%%%%%%%%%%%%%%%%%%%%%%%%%%%%%%%%%%%%%%%%%%%%
% Discussion
%%%%%%%%%%%%%%%%%%%%%%%%%%%%%%%%%%%%%%%%%%%%%%%%%%%%%%%%%%%%%%%%%%%%%%%%%%%%%%%%%%%%%%%%%%%%%%%%%%%%%%%%%%%%%%%%%%%%%%%%%%%%%%

\section{Evolution of criticality in the elastic regime}\label{chap:discussion}
In this chapter, the results in the elastic regime are presented.
Specifically, to discuss the issues of criticality and universality independently,
we separately present the results of the ensembles of only the initial avalanche events of different samples and of the avalanches collected over the entire elastic regime $\gamma\in[0,0.02]$~\cite{Shang2020}.

%%%%%%%%%%%%%%%%
%%%%%%%%%%%%%%%%
%%%%%%%%%%%%%%%%
\subsection{Results for unperturbed systems}\label{sec:first}
\begin{figure}
  \includegraphics[width=\linewidth]{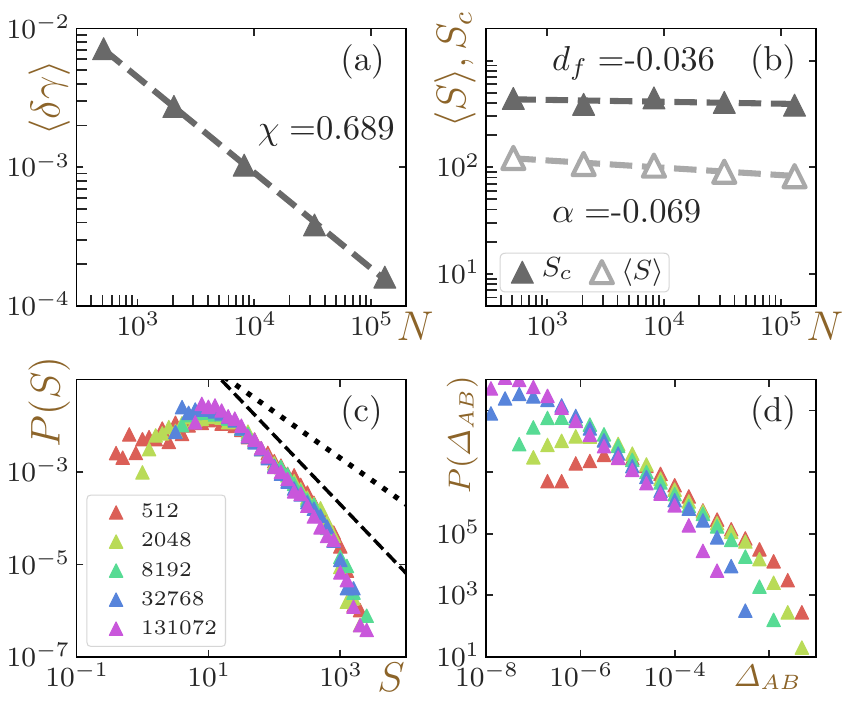}
\caption{
Statistics of the first event ensemble.
System size dependence of (a) $\langle \delta\gamma\rangle$,
(b) $\langle S \rangle$ and $S_c$.
The markers indicate the numerical results, and the lines are a power-law fitting.
(c) PDFs of the avalanche sizes $S$.
Different colors are used for different system sizes, as shown in the
legend.
The dashed line depicts the results of power-law behavior with the exponent in the steady state,
$\tau\approx 1.493$, and the dotted line denotes $\tau=1.0$.
(d) PDF of the MSD $\Delta_{AB}$.
The meanings of the colors are the same as in (c).
    \label{fig:first_4}}
\end{figure}

%%%%%%%%%%%%%%%%
%%%%%%%%%%%%%%%%
%%%%%%%%%%%%%%%%
First, we present the results of the ensembles of only initial
avalanche events, which should most strongly reflect the features of
unperturbed systems.
For each system size, we prepared 4000 independent samples
and applied simple shear in an AQS manner until we encountered
avalanches.
To judge whether the first obtained avalanche event was a
precursor or a mainshock, we detected the first two events.
The important statistical information is summarized in
Fig.~\ref{fig:first_4}.

\emph{Independent exponents ---.}
{The system size $N$ dependence of $\langle\delta\gamma\rangle$ and
$S_c$, which yield the independent exponents $\chi$ and $d_f$,
are shown in Fig.~\ref{fig:first_4}(a,b).}
Although $\langle\delta\gamma\rangle$ exhibits power-law system size
dependence, as in the steady state, $S_c$ (and $\langle S\rangle$) appears rather constant.
This result is consistent with the findings reported in ref.~\cite{Karmakar2010a} and means that
criticality is absent in unperturbed systems.

Note that Lin and coworkers have theoretically shown
that the pseudogap exponent of an unperturbed system should be $\theta=0.5$ universally~\cite{Lin2016}.
This means that $\chi$ should be $2/3$, and
our numerical result $\chi=0.689$ is reasonably close to this theoretical prediction.

%------------------------------------------------------------------------------------------
%%% Figure: 4 panels first event ensemble
%------------------------------------------------------------------------------------------
%------------------------------------------------------------------------------------------

%%%
\emph{Avalanche size distribution ---.}
Since the statistics are system size-independent,
the PDFs of the avalanche sizes of different system sizes
are almost identical without any scaling (see Fig.~\ref{fig:first_4}(c)).
Interestingly, the
PDFs still show broad power-law-like shapes.
However, because of the absence of criticality,
we cannot draw any absolute conclusion as to
whether they do indeed follow a power law,
although their apparent \emph{slopes} seem consistent
with the value in the steady state,
$\tau\approx 1.493$.
Nevertheless, we can safely conclude that their apparent \emph{slopes} are much larger than
$\tau=1.0$, which is the prediction of the theory in ref.~\cite{Franz2017}.
We also note that the PDFs do not show bumps in the large size regime,
unlike the steady state results.

\begin{figure}
  \includegraphics[width=0.7\linewidth]{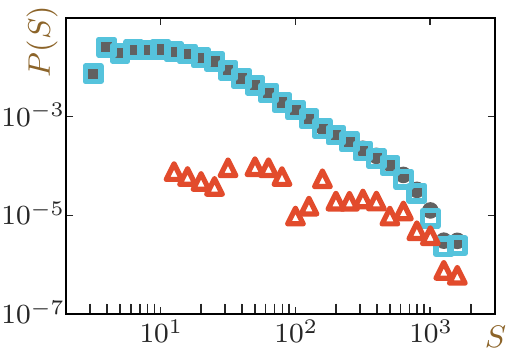}
\caption{
Decomposition of the PDF of the avalanche sizes of the first event ensemble
into the contributions from precursors and mainshocks.
The results for systems with $N=32768$ are shown.
The meanings of the markers are the same as in
Fig.~\ref{fig:steady_pre_main}.
    \label{fig:first_pre_main}
}
\end{figure}

%%%
\emph{Decomposition of avalanche size distribution ---.}
In Fig.~\ref{fig:first_pre_main}, we demonstrate that the PDF of the
first event ensemble can also be decomposed into contributions
from the precursors and mainshocks.
In this case, the mainshocks are not well developed and are
completely obscured by the precursors.
This is why there is no bump in the PDFs.

%------------------------------------------------------------------------------------------
%%% Figure: Pre/Main separation of Avlanche/MSD distribution (first)
%------------------------------------------------------------------------------------------
%------------------------------------------------------------------------------------------

%%%
\emph{Mean square displacements ---.}
{The theory in ref.~\cite{Franz2017} discussed} the relation
between the energy landscape in the Gardner phase and the statistics
of avalanches induced by a very weak perturbation
(see Appendix~\ref{ap:Gardner} for a brief summary of the Gardner transition and this theory).
Since this theory is based on the replica method, an
\emph{equilibrium} statistical mechanics theory, the first event
ensemble that represents the unperturbed system is expected to correspond well to the theoretical situation, although thus far, the avalanches during a small but finite range of strains have been treated as numerical counterparts~\cite{Franz2017,Shang2020}.
Here, we quantify the degree of the similarity to the Gardner phase and show that the
presupposition of the full breakage of the replica symmetry
in the theory is not satisfied in the case of the first event
ensemble in our LJ glass system.
This is likely to be one of
the reasons why the results are inconsistent with the theoretical
predictions~\cite{Franz2017}, such as those for criticality and the value of
$\tau$.

The mean squared displacement (MSD) between
different realizations of configurations, $\Delta_{AB}\equiv
\frac{1}{N}\sum_i^{N}(\boldsymbol{r}_i^A-\boldsymbol{r}_i^B)^2$, is a
quantitative measure of the dissimilarity between two
configurations,
and its PDF can be used as an order parameter for the Gardner phase\footnote{There are several quantitative measures of the
similarity or dissimilarity between two configurations. We employed the MSD
because it is defined without any parameters. }.
Here, $\boldsymbol{r}_i^s$ is the position of particle $i$ in the
configuration $s\in {A, B}$,
where states A and B are again used for the states before and after
an avalanche
event, respectively.
According to the mean-field replica theory, in the Gardner phase,
the system shows a very wide PDF of the MSD, and
importantly, $\Delta_{AB}$ possesses a
system-spanning nature reflecting the diverging correlation length~\cite{Charbonneau2015,Scalliet2017}.
Since the MSD $\Delta_{AB}$ is defined as an intensive variable, the
maximum value of the MSD, which
corresponds to the system-spanning structure,
is expected to be the same across
different system sizes if the systems are in the Gardner
phase~\cite{Parisi2020}.

In Fig.~\ref{fig:first_4}(d), the PDFs of the MSDs of different system sizes
are plotted.
Both the maximum and minimum edges of the PDFs shift in accordance with the
change in the system size.
This behavior is qualitatively different from what we expect
for the full PDF of the MSD in the Gardner phase, where the maximum edge
should be constant regardless of
the system size.
Rather, the PDFs are consistent with
those of avalanche sizes: they show broad
distributions but are not system spanning.
The localized (non-system-spanning) tendency can be more explicitly
quantified by the total squared displacement (TSD)
$\Sigma\Delta_{AB}\equiv N\Delta_{AB}$, which carries information regarding the
geometrical size of avalanches. The
PDFs of the TSDs of different system sizes nearly overlap each other
without any scaling,
as is the case for the PDFs of the avalanche sizes (see Appendix~\ref{ap:TSD}).
Therefore, we conclude that the unperturbed systems do not share the same system-spanning vulnerability that is expected for the Gardner phase.
Recently, in refs.~\cite{Scalliet2017,Hicks2018,Seoane2018}
the existence of the Gardner phase in
physical, finite-dimensional soft-potential systems at rest
(unperturbed systems) has been denied.
Our results are consistent with these studies.

%%%%%%%%%%%%%%%%
%%%%%%%%%%%%%%%%
%%%%%%%%%%%%%%%%
\subsection{Results of weakly perturbed systems}\label{sec:transient}
\begin{figure}
  \includegraphics[width=\linewidth]{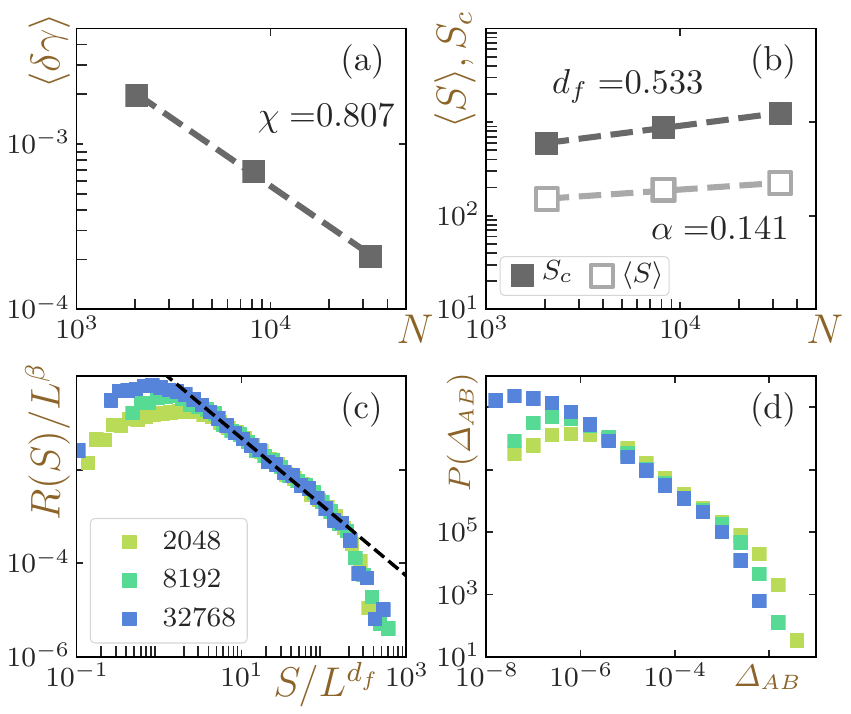}
\caption{
Statistics of avalanches in the elastic regime $\gamma\in [0, 0.02]$.
System size dependence of (a) $\langle \delta\gamma\rangle$,
(b) $\langle S \rangle$ and $S_c$.
The markers represent the numerical results, and the lines are the power-law fitting.
(c) Scaled PDFs of the avalanche sizes $S$ per unit strain.
Different colors are used for different system sizes, as shown in the
legend.
The dashed line indicates power-law behavior, with the exponent
estimated by Eq.~\ref{eq:robust_scaling},
$\tau\approx 1.471$.
(d) PDF of the MSD $P(\Delta_{AB})$.
The meanings of the colors are the same as in (c).
    \label{fig:transient_4}}
\end{figure}

%%%%%%%%%%%%%%%%
%%%%%%%%%%%%%%%%
%%%%%%%%%%%%%%%%
We conduct the same analysis in the weakly perturbed elastic regime,
$\gamma\in [0,0.02]$\cite{Shang2020}.
We performed simulations of 600, 160, and 50 samples for systems with $N=2048, 8192,$ and
$32768$, respectively.
These sample numbers are chosen to guarantee more than 4000 events for each system
size.
The important statistical information is summarized
in Fig.~\ref{fig:transient_4}.

%%%
\emph{Independent exponents ---.}
In this case, all $\langle \delta\gamma\rangle$, $\langle S\rangle$
and $S_c$ show a power-law dependence on the system size $N$ (see Figs.~\ref{fig:transient_4}(a,b)).
The value of $\chi$ is larger than that in the steady state
because of the well-known non-monotonicity of the pseudogap exponent
$\theta$,
which was first predicted theoretically~\cite{Lin2016} and then
numerically confirmed~\cite{Ozawa2018,Shang2020}.
Nonzero values of $\alpha$ and $d_f$ indicate criticality in the elastic regime, in accordance with ref.~\cite{Shang2020}.
However, the estimated values of $\chi$ and $\alpha$ do not meet
Eq.~\ref{eq:alpha_relation}, $\alpha=1-\chi$, since the stationarity
condition Eq.~\ref{eq:ave_S2} is trivially unsatisfied in this regime.
We note that the same degree of discrepancy between $\alpha$ and
$1-\chi$ in the elastic regime was reported in ref.~\cite{Shang2020}
\footnote{This seems to have nothing to do with the fact that
$\theta$ changes abruptly and non-monotonically in the vicinity of
$\gamma=0$ because, even if we restrict ourselves to the strain range
$\gamma\in[0.005, 0.02]$
in which $\theta$ temporarily becomes constant, as in ref.~\cite{Shang2020}, we observe the same results
semi-quantitatively. }.
We also mention that the fractal dimension $d_f$ in the elastic regime
is much less than that in the steady state.

%%%
\emph{Avalanche size distribution ---.}
The failure of Eq.~\ref{eq:ave_S2} means that Eq.~\ref{eq:Lin_scaling}
is not applicable in this regime
either\footnote{
Eq.~\ref{eq:Lin_scaling} is expected to be valid even in the elastic
regime if we conduct the measurement at a fixed stress level as in ref.~\cite{Lin2015}.
}.
However, by comparing Eq.~\ref{eq:ave_S1} and the definition of the
exponent $\alpha$, we can derive another form of the scaling relation:
\begin{align}
  \tau=2-\alpha\frac{d}{d_f}.\label{eq:robust_scaling}
\end{align}
This relation is robustly usable in the elastic regime.
The estimation of $\tau$ with this new relation is $\tau\approx 1.471$,
and it again describes the small-size regime of the PDFs of
the avalanche sizes well (see Fig.~\ref{fig:transient_4}(c)).
Moreover, we stress that this value is
very close to the value in the steady state and the mean-field prediction.
This result is in
agreement with the experimental observations~\cite{Antonaglia2014,Denisov2016}.

%------------------------------------------------------------------------------------------
%%% Figure: 4 panels transient 1
%------------------------------------------------------------------------------------------
%------------------------------------------------------------------------------------------

\begin{figure}
  \includegraphics[width=0.7\linewidth]{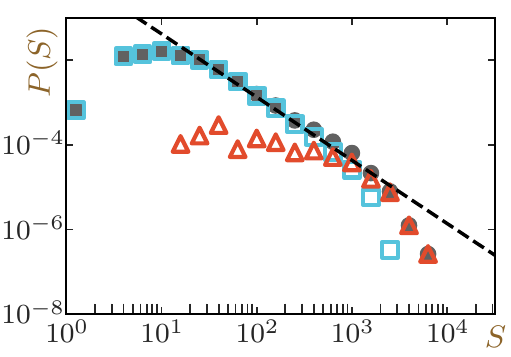}
\caption{
Decomposition of the PDF of the avalanche sizes in
the elastic regime
into the contributions from precursors and mainshocks.
The results for systems with $N=32768$ are shown.
The meanings of the markers are the same as in Fig.~\ref{fig:steady_pre_main}.
The dashed line depicts the power-law behavior, with the exponent shown in
Fig.~\ref{fig:steady_ava}. \label{fig:transient_pre_main}}
\end{figure}

\emph{Decomposition of avalanche size distribution ---.}
The decomposition of the PDF of the avalanche sizes
into the contribution from precursors and
mainshocks again provide much information (Fig.~\ref{fig:transient_pre_main}).
The PDF of the precursors shows normal power-law behavior
with a cutoff, as is the case in the steady state, and moreover, the exponent seems
to remain the same.
In the elastic regime, unlike the case of the first event ensembles,
the mainshocks form a small bump in the large-size regime.

Note that, as presented in Appendix~\ref{ap:comparison2},
since mainshocks come into play in this regime,
we can fit the data in the crossover regime (which is close to the bump) directly by a power-law curve.
The obtained value of the avalanche exponent $\tau^\prime$
is quantitatively consistent
with that in ref.~\cite{Shang2020}, $\tau^\prime\approx 1.0$,
where $\tau^\prime$ is the avalanche exponent estimated by
potential drops.

%------------------------------------------------------------------------------------------
%%% Figure: Pre/Main separation of Avlanche/MSD distribution (transeint)
%------------------------------------------------------------------------------------------
%------------------------------------------------------------------------------------------

%%%
\emph{Mean square displacements ---.}
The characteristics of the criticality in this regime can be
quantified by the PDF of the MSD.
As shown in Fig.~\ref{fig:transient_4}(d), the maximum values of the PDFs of the MSDs of different system sizes still
exhibit small discrepancies with each other.
Still, reflecting criticality, the
PDFs of the TSD are system size-dependent (Appendix~\ref{ap:TSD}).
We stress that if we analyze the displacement field during the
avalanche event with the largest avalanche size in the elastic regime,
the mobile particles are indeed system spanning.
However, the structure of the cluster of mobile particles is less packed than
that in the steady state, manifesting its smaller fractal dimension.

\subsection{MSD in the steady state}\label{sec:steady_MSD}
\begin{figure}
  \includegraphics[width=\linewidth]{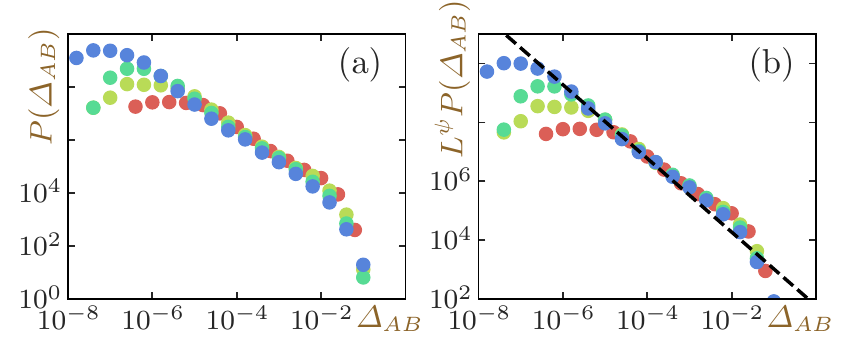}
\caption{
PDFs of the MSD $\Delta_{AB}$ in the steady state
(a) without scaling, and (b) with finite-size scaling, where $\psi=0.3$.
The dashed line in (b) is a guide for the eye to stress the power law
behavior $P(\Delta_{AB})\sim \Delta_{AB}^{-\tau^*}$ with
$\tau^*=1.25$.
The colors of the markers have the same meaning as in Fig.~\ref{fig:steady_ava}.
    \label{fig:steady_MSD}}
\end{figure}

Finally, we demonstrate that avalanches in the steady state are indeed
fully system spanning.
We plot the PDFs of the MSD during avalanche events $\Delta_{AB}$ in the
steady state
in Fig.~\ref{fig:steady_MSD}(a).
The PDFs of the MSD all have broad distributions,
and in particular, the maximum values of different system sizes are the
same.
This is exactly what we expect for the Gardner phase, as presented
above in Sec.~\ref{sec:first}.
Therefore, the criticality of the avalanches in binary LJ glass in the
steady state
is remarkably similar to that in the Gardner phase with respect to the
spatial structure,
although we are still not sure how tightly we can
connect these two concepts because of the lack of a theoretical
description.
Jin and coworkers~\cite{Jin2018}
have reported the existence of
a shear-induced Gardner transition in a hard sphere system.
The extension of their work to softer potentials would be one
promising way to test whether our findings have similar
characteristics to theirs.
Since the MSD $\Delta_{AB}$ is a particle-averaged variable,
events with the same TSD result in smaller values of $\Delta_{AB}$
in larger systems.
This characteristic leads to a difference in the
range of the PDFs depending on the system size: the larger
the system becomes, the wider the range becomes (the
smaller the smallest MSD becomes).
{Reflecting this difference in the range, the PDFs of MSDs $P(\Delta_{AB})$
for different
system sizes can be collapsed by scaling as $L^\psi P(\Delta_{AB})$,
as shown in Fig.~\ref{fig:steady_MSD}(b).}
In agreement with the PDFs
of avalanche sizes, the power-law regime can be seen in a
small value regime, and there is a bump in the large value
regime. We mention that the precursor/mainshock
decomposition is also valid for PDFs of the MSD, and again the
bump is composed only of mainshocks (figure not shown).

%------------------------------------------------------------------------------------------
%%% Figure: MSD distribution and its finite size scaling
%------------------------------------------------------------------------------------------
%------------------------------------------------------------------------------------------

\subsection{Unified view of avalanche criticality}
All the results presented thus far provide a unified understanding of 
avalanche criticality and universality in the sheared LJ glass system
throughout the whole strain regime.
% criticality
While the first event ensemble that represents the
unperturbed system is off-critical,
criticality emerges
in both the elastic regime and the steady state.
However, the fractal dimension of the avalanches indicates
the quantitative difference between these two strain regimes.
In particular, the elastic regime is described by a smaller value of $d_f$.
In this sense, we conclude that criticality gradually grows as 
shear is applied and becomes fully developed in the steady state, where
the fractal dimension is saturated.
These observations are all consistent with previous
works in the literature~\cite{Karmakar2010,Lin2015,Seoane2018,Hicks2018,Shang2020}.

% universality
{ Once
the system becomes critical, the critical exponent
remains constant, regardless of the amount of applied strain, 
and importantly, this value is consistent with the mean-field
prediction $\tau_{\rm MF}=1.5$.}
The universality of the critical exponent in the elastic regime is
then at odds with the
theoretical prediction made in ref.~\cite{Franz2017}.
This might be partly because the unperturbed system is not in
the Gardner phase.
Finally, we repeatedly emphasize that the universal values of $\tau$ in each regime in our work are different from those in
all previous numerical studies\cite{Salerno2012,Salerno2013,Zhang2017,Ozawa2018,Saitoh2019,Shang2020}.
We consider that non-universal values in previous studies might have been obtained
because the crossover regime resulting from the finite-size effects has been
analyzed.

%%%%%%%%%%%%%%%%%%%%%%%%%%%%%%%%%%%%%%%%%%%%%%%%%%%%%%%%%%%%%%%%%%%%%%%%%%%%%%%%%%%%%%%%%%%%%%%%%%%%%%%%%%%%%%%%%%%%%%%%%%%%%%
% Conclusion
%%%%%%%%%%%%%%%%%%%%%%%%%%%%%%%%%%%%%%%%%%%%%%%%%%%%%%%%%%%%%%%%%%%%%%%%%%%%%%%%%%%%%%%%%%%%%%%%%%%%%%%%%%%%%%%%%%%%%%%%%%%%%%
\section{Conclusion and overview}\label{chap:conclusion}
Here, we conducted a thorough investigation of avalanche
criticality in a sheared binary LJ glass system by means of atomistic simulations.
In particular, by ruling out the ambiguity and arbitrariness that have
slipped into measurements in previous studies, we first showed that
the critical avalanche exponent in the steady state coincides with the mean-field prediction~\cite{Dahmen2009}.
Our results simultaneously suggest that the scaling function of the avalanche size distribution has a nontrivial bumpy shape.
We noticed that there are two qualitatively different avalanche events,
and this binariness explains the physical origin of the strange bump in the scaling function.
Furthermore, we demonstrated that this bump is likely to be the cause of the nonuniversal results for the critical exponent $\tau$ obtained in previous studies (Appendix~\ref{ap:comparison1}).

To investigate the change in criticality and
universality due to applied shear,
we conducted the same high-precision measurements of avalanche statistics of the first event ensemble, which reflect the properties of the unperturbed system and avalanches in the elastic regime.
As a result, we confirmed that the first event ensemble does not exhibit any system-size dependence and thus lacks criticality.
This consequence dovetails with the result in ref.~\cite{Karmakar2010a}.
Avalanches in the elastic regime, on the other hand, do display
criticality, in accordance with refs.~\cite{Lin2015,Shang2020}, and again,
the exponent of the power-law part is very close to the mean-field
value universally.
The value of the critical exponent is in accordance with recent experiments conducted in
the elastic regime~\cite{Antonaglia2014,Denisov2016}.
The criticality in the elastic regime is different from that in the
steady state and is characterized by a much smaller value of the fractal
dimension $d_f$.
We believe that our results provide a unified picture of
avalanche criticality in deformed glasses, for which confusing and seemingly conflicting results have been reported thus far: 
criticality itself develops along with applied strain, with the exponent
of the power-law part remaining constant.
In particular, the change in criticality is quantitatively encoded in
the fractal dimension $d_f$, which takes the value of zero in the
off-critical unperturbed state and saturates in the steady state.

We employed configurations that are not in the Gardner phase as the initial state
in this work, and thus,
the starting point itself differs from the theory in ref.~\cite{Franz2017},
where a system in the Gardner phase is considered the
initial state. If we find the Gardner phase in physical-dimensional
amorphous solid systems with soft potentials,
it would be very important and meaningful
to conduct the same analyses using the configuration in the
Gardner phase as the initial state.

Since we applied shear in an AQS way,
the dynamical information could not be accessed.
It would also be very important to conduct simulations with a finite-rate shear and investigate the dynamical information, such as the
avalanche duration, the
avalanche shape and the power spectrum of the stress-drop rate time series.

\begin{acknowledgments}

We thank H. Yoshino, M. Ozawa and H. Ikeda for useful discussions.
This work was financially supported by KAKENHI grants
(nos.  18H05225, 19H01812, 19K14670, 20H01868, 20H00128, 20K14436 and 20J00802) and partially supported by the Asahi Glass Foundation.

\end{acknowledgments}

\appendix

\section{Comparison with previous works}\label{ap:comparison}
\subsection{Avalanche exponent in the steady state}\label{ap:comparison1}
In this appendix, we discuss the cause of the discrepancies between our
result for the avalanche exponent $\tau\approx 1.493$ in the steady
state and those in
previous works with atomistic simulations, $1.15\le \tau \le 1.3$~\cite{Salerno2012,Salerno2013,Zhang2017}.
To this end, we measured $\tau$ by the same method as in the
previous works: a direct fitting to the PDFs of avalanche sizes.
In particular, we utilized only the data in the large size regime, where
the cut-off size $S_c$ resides.
As shown in Fig.~\ref{fig:previous}(a),
the resulting exponent $\tau=1.19$ is located in the middle of the values
reported in previous works.
Thus, we consider that the cause of the variation in the value of
$\tau$ might result from the fact that the non-universal part, which can depend on the
details of the systems, has been fitted.

\subsection{Avalanche exponent in the elastic
regime}\label{ap:comparison2}
Although we used the stress drops for the definition of the avalanche
sizes in the main text,
the avalanche sizes can also be defined based on the potential energy
drops, as follows:
\begin{align}
    S^\prime_i \equiv \Delta E_{ABi}\equiv E_{A}(\gamma_{Ci})-E_{B}(\gamma_{Ci}),\label{eq:ava_pe}
  \end{align}
where $E_s(\gamma)$ is the potential energy of state $s\in A,B$ at
strain $\gamma$.
We use the prime symbol to express the variables of the avalanches defined
by Eq.~\ref{eq:ava_pe}.

\begin{figure}
  \includegraphics[width=\linewidth]{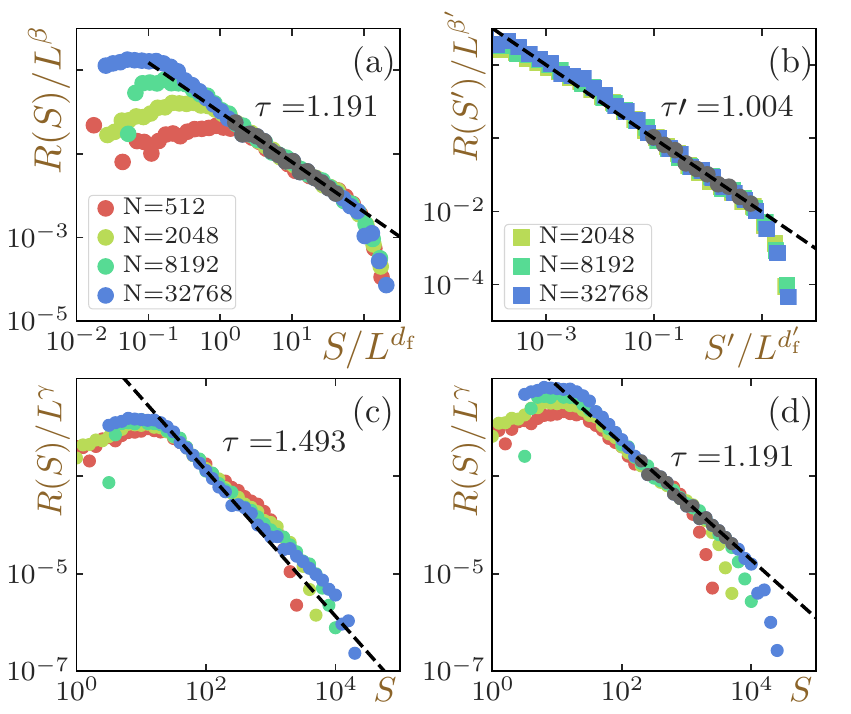}
\caption{
(a) Scaled PDFs of the avalanche sizes defined based on the stress drops $S$ in the steady
state.
(b) Scaled PDFs of the avalanche sizes defined based on the potential
energy drops $S^\prime$ in the elastic regime.
(c) Scaled PDFs of the avalanche sizes with the exponent $\gamma$
estimated by $\tau=1.493$, the value obtained in the main text.
(d) Scaled PDFs of the avalanche sizes with the exponent $\gamma$
estimated by $\tau=1.19$, the value obtained in (a).
In all panels, different colors and symbols indicate different system sizes, as shown in the
legends in (a, b).
The dashed lines in (a, b, d) depict the results of direct fitting
by using the data points highlighted in gray.
The estimated values of the avalanche exponents $\tau$ and
$\tau^\prime$ are also shown.
     \label{fig:previous}
}
\end{figure}

{We also conducted the same direct fitting to the PDF of
$S^\prime$, avalanche sizes defined based on the potential energy
drops,
in the elastic regime.}
The fitting result is $\tau^\prime=1.004$ and is perfectly consistent with
the result in ref.~\cite{Shang2020}, as shown in
Fig.~\ref{fig:previous}(b).

%------------------------------------------------------------------------------------------
%%% Figure: Exponents in the previous study
%------------------------------------------------------------------------------------------
%------------------------------------------------------------------------------------------

\subsection{Scaling collapse}\label{ap:comparison3}
We further demonstrate that the collapse of the data of
different sizes by a scaling law is too robust, and thus,
unfortunately, it is not reliable enough to guarantee the correctness of the results.
In refs.~\cite{Salerno2012,Salerno2013,Zhang2017},
the estimated value of $\tau$ is validated by the following scaling
law:
\begin{align}
  \gamma=\beta+d_f\tau\label{apeq:scaling},
\end{align}
where another scaling exponent
$\gamma$~\footnote{Although we already used the letter $\gamma$ to refer
to the applied strain, we name this exponent $\gamma$
following the definition of the original papers~\cite{Salerno2012} to avoid
any confusion.} is introduced:
by scaling by $L^\gamma$,
the \emph{power-law parts} of the PDFs of different system sizes can be
collapsed, as shown in Fig.~\ref{fig:previous}(c).
Here, we estimated the value of $\gamma$ by using only $d_f$ and
$\chi$, as $\gamma=\beta+d_f\tau=d\chi$, as is the case for the other
exponents.
Note that the authors of refs.~\cite{Salerno2012,Salerno2013,Zhang2017} also confirmed that Eq.~\ref{eq:beta_scaling1} holds
for the obtained exponents.

What if we try the same scaling collapse with $\tau=1.19$ that we
obtained by the direct fitting to the data in
Fig.~\ref{fig:previous}(a)?
The results are shown in Fig.~\ref{fig:previous}(d) (in this
case, we use Eq.~\ref{apeq:scaling} to obtain the exponent $\gamma$).
As seen here, the \emph{power-law parts} of different system sizes are
collapsed again, even with a different value of $\tau$.
This means that Eq.~\ref{apeq:scaling} can be satisfied, unexpectedly, too
robustly with multiple values of $\tau$, and
the successful collapse by Eq.~\ref{apeq:scaling}
alone is not enough evidence for the validity of the obtained value of $\tau$.

\section{Are mainshocks induced by the inertial effect?}\label{ap:bumps}
Several studies have reported that the introduction of inertia
can induce bumps in the
PDFs of avalanche sizes~\cite{Salerno2012,Salerno2013,Karimi2017}, which are similar to our mainshocks.
{We discuss the differences among our bumps and those of other studies}.

%\subsection{Inertial effect}
In ref.~\cite{Karimi2017},
Karimi and coworkers investigated the effect of inertia on the
PDFs of avalanche sizes by numerical simulation of a finite-element-based
elastoplastic model.
They reported that as the effect of inertia
becomes stronger, the PDF of the avalanche sizes begins to exhibit a
characteristic bump in the large size regime.
Since we employed the FIRE algorithm in this work, which can introduce
an inertial effect during the energy minimization process,
it is possible that our mainshocks share the same origin as the bump
reported in ref.~\cite{Karimi2017}.

\begin{figure}
  \includegraphics[width=0.6\linewidth]{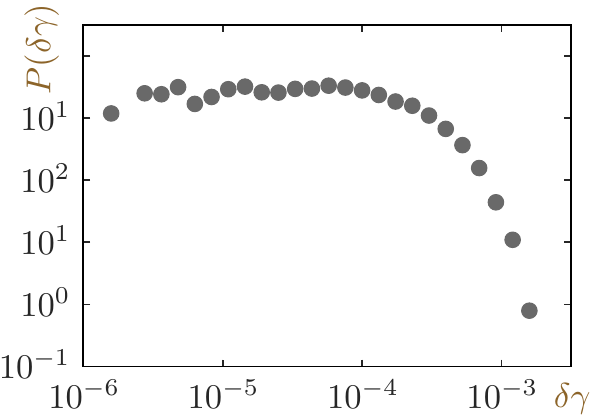}
\caption{
PDF of $\delta \gamma$, or the local stability $x_{\rm min}$ in
our setup.
The steady-state result of a system with $N=32768$ is shown.
    \label{fig:xmin}}
\end{figure}

{
  According to Karimi {\it et al.}, in the case of their elastoplastic model, the PDF
of the minimum of the local stability $x_{\rm min}$ shows
a bimodal nature when inertia takes effect,
and the peak in the large value of $x_{\rm min}$ is responsible for
the bump in the PDF of the avalanche sizes.
To check whether our bump has the same physical origin,
we also measured the PDF of $\delta\gamma$, which
represents the minimum of the local stability in our setup, as discussed
in Sec.~\ref{sec:scaling}.}
As shown in Fig.~\ref{fig:xmin},
in our case, the PDF of $\delta\gamma$ does not exhibit any salient
peaks in the large value regime.
This means that the bump observed in our work has a qualitatively different
physical origin from that in ref.~\cite{Karimi2017}.
We note that the unimodal shape of $P(\delta\gamma)$ is consistent with
refs.~\cite{Karmakar2010a,Zhang2017}.

Similar inertial effects have also been found in the atomistic simulations in
refs.~\cite{Salerno2012,Salerno2013}.
Their results are qualitatively similar to those in our study
(the bumps in their PDFs exhibit the same scaling exponents as
the ones for the power-law regime).
However, the avalanche exponents obtained in the power-law regime are very different ($\tau=1.0$ and
$1.25$ in their inertial cases).
Therefore, we conclude that the bump in our scaling function is
different from those in refs.~\cite{Salerno2012,Salerno2013}, and thus,
our mainshocks are not due to the inertial effect possibly
caused by the FIRE algorithm.
Indeed, similar bumpy shapes have also been observed in studies
where inertialess energy minimization protocols were
employed~\cite{Lin2014,Lerner2014a,Zhang2017}.

%------------------------------------------------------------------------------------------
%%% Figure: PDF of the local stability
%------------------------------------------------------------------------------------------
%------------------------------------------------------------------------------------------

\section{Validation of the strain resolution}\label{ap:validation}
\begin{figure}
  \includegraphics[width=\linewidth]{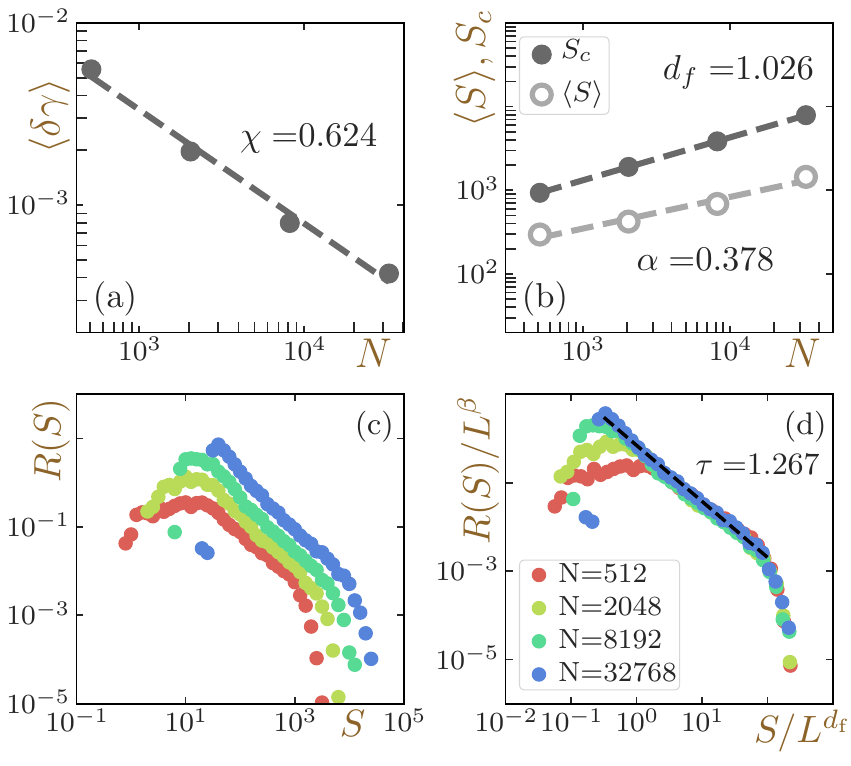}
\caption{
Statistics of avalanches in the steady state with a fixed crude strain
resolution of $\Delta\gamma=1\times 10^{-5}$.
System size dependence of (a) $\langle \delta\gamma\rangle$,
(b) $\langle S \rangle$ and $S_c$.
The markers represent the numerical results, and the lines are the power law fitting.
(c) Unit strain PDFs of the avalanche size $S$. (d) Scaled unit
strain PDFs of the avalanche size $S$.
The dashed line in (d) depicts the power-law behavior predicted by
the scaling law Eq.~\ref{eq:Lin_scaling} with
values of $\chi$ and $d_f$ shown in (a,b), $\tau\approx 1.276$.
In both (c) and (d), different colors indicate different system sizes, as shown in the
legend in (d).
    \label{fig:validation1}}
\end{figure}

In this appendix, we explain how the strain resolution $\Delta\gamma$
was determined.
First, to provide intuition into the importance of the tuning of
$\Delta\gamma$,
we show the results with a fixed crude value of $\Delta\gamma=1\times
10^{-5}$ in Fig.~\ref{fig:validation1}.
Here, the statistical information in the steady state is shown.
As shown in Fig.~\ref{fig:validation1}(a,b), the $N$ dependence of
$\langle\gamma\rangle$ and $\langle S\rangle$ obviously do not have power-law
shapes anymore (the clear deviation of the data of $N=32768$ can be recognized).
If we turn our attention to the PDF of the avalanche sizes (Fig.~\ref{fig:validation1}(c)),
the data of the largest system size, $N=32768$, do not reach the peak
in the small size regime due to the lack of resolution.
If we still attempt to fit the $N$-dependence of $\langle\gamma\rangle$ and
$S_C$ to power-law curves and \emph{estimate} the exponents $\chi$,
$d_f$ and then $\tau$
from these data,
we obtain $\tau\approx 1.267$.
Note that, as shown in Fig.~\ref{fig:validation1}(d),
this exponent does not seem to be inconsistent with the entire curve,
and it is very difficult to
tell that the result is incorrect if one looks only at the PDF of the
avalanche sizes and not at the $N$ dependence of $\langle \delta\gamma\rangle$.
To summarize, the lack of resolution can lead to the deviation of the $N$ dependence of
$\langle\delta\gamma\rangle$ and $\langle S\rangle$ (and presumably $S_c$ as
well) from the expected power-law behavior.
Moreover, the PDF of the avalanche sizes is truncated from the small size
regime where the intrinsic power-law part resides.
We carefully tuned the strain resolution $\Delta \gamma$ so that none of
these problems appear.

\begin{figure}
  \includegraphics[width=\linewidth]{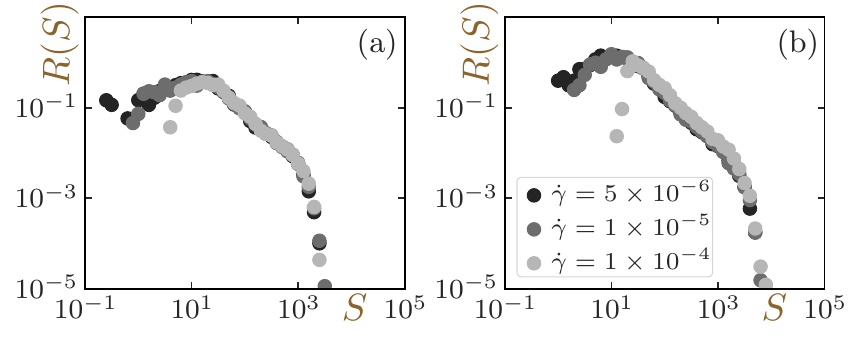}
\caption{
Comparison of the PDFs of the avalanche sizes $S$ with different strain
resolutions $\Delta\gamma$.
The results in the steady state are shown.
(a) Results for $N=512$. (b) N=2048.
Different colors represent different strain resolutions
$\Delta\gamma$, as shown in the legend in (b).
    \label{fig:validation3}}
\end{figure}

We further note that to tune the resolution according to the
procedure presented above, we need reliable data for small systems as a
reference.
For this purpose, we conducted simulations for small sizes,
$N=512, 2048$, with different resolutions, $\Delta\gamma=1\times
10^{-4}, 1\times 10^{-5}, 5\times 10^{-6}$ and confirmed that the PDF
converges for $\Delta\gamma\le 1\times 10^{-5}$
(Fig.~\ref{fig:validation3}).
This is why we employed the value $\Delta\gamma=5\times 10^{-6}$ for small systems.

%------------------------------------------------------------------------------------------
%%% Figure: Validation1
%------------------------------------------------------------------------------------------
%------------------------------------------------------------------------------------------

%------------------------------------------------------------------------------------------
%%% Figure: Validation3
%------------------------------------------------------------------------------------------
%------------------------------------------------------------------------------------------

\section{Marginal stability}\label{ap:ms}
In the last decade,
the relation between avalanches in sheared
amorphous solids and the marginal stability,
which is now expected to be a distinguishing universal feature of
amorphous solids, has been discussed~\cite{Muller2015}.
In this appendix, we briefly summarize two concepts related to our work.

\subsection{Pseudogap exponent}\label{ap:pseudogap}
To provide insight into the first concept, the pseudogap exponent,
we first introduce a so-called elastoplastic model.
In this model, an amorphous solid system is assumed to be an assembly of
mesoscale sites.
Each site corresponds to a coarse-grained description of a group of
particles and has its own local strain $\gamma_i$, local stress
$\sigma_i$ and local yield stress $\sigma_i^{\rm th}$.
When the local stress of a site exceeds the corresponding local yield stress,
the site will yield locally and give rise to a plastic strain.
Such a local yielding is associated with an STZ and thus affects other
sites' local stresses, causing avalanches.
The marginal stability can be reflected by, in this context, a
pseudogap in the PDF of the local
distance to yielding $x_i\equiv \sigma_i^{\rm th}-\sigma_i$\cite{Lin2014};
the PDF of $x_i$ obeys a power law, $P(x)\sim x^\theta$, {in the limit of $x\to 0$.}
This means that the system is marginally stable against external fields.
The existence of the pseudogap in $P(x)$ has been confirmed
numerically by means of both atomistic simulations and the elastoplastic model,
in both two and three dimensions~\cite{Karmakar2010a,Lin2014}.
Lin and coworkers
derived a scaling relation among the avalanche exponent $\tau$,
the pseudogap exponent $\theta$ and
the fractal dimension $d_f$ of the spatial structures of
avalanches\cite{Lin2014}, as presented in the main text (see Eq.~\ref{eq:Lin_scaling}).
They validated the scaling relations by numerical calculations
with an elastoplastic model.

{\subsection{Gardner transition}\label{ap:Gardner}
Another distinctive feature of the marginal stability has been
predicted by infinite-dimensional mean-field replica
theory\cite{Charbonneau2014}.
This theory states that when a glassy system crosses a specific border
in the parameter space, it experiences full-replica
symmetry breakage, and
there abruptly emerge infinitesimally different (almost identical)
metastable states.
This special glassy phase after the transition is distinguished from
normal glassy states and is called the Gardner phase.
The nature of the Gardner phase is reflected in
the infinite hierarchy of metabasins in the energy
landscape, and
several works have confirmed that the Gardner phase can be observed
even in finite physical dimensions in hard sphere
systems\cite{Berthier2016a,Jin2018}.
However, the parameter space for the Gardner phase is severely limited,
and, at least thus far, no work has detected the Gardner phase in
a system with softer potentials, such as LJ or the inverse
power law, in physical
dimensions~\cite{Scalliet2017,Hicks2018,Seoane2018}.
Thus, the Gardner aspect
of the marginal stability is currently a matter of very active debate~\cite{Berthier2019}.

Franz and Spigler discussed the relation between avalanche
statistics and the Gardner phase in an amorphous solid system with a genuinely short-ranged potential in which the jamming criticality also plays a major role~\cite{Franz2017}.
They first formulated, by the replica method,
the hierarchical structure of the energy landscape in the Gardner phase.
They then treated plastic events under external shear as
transitions between metabasins with perturbations induced by shear
and showed that, corresponding to {the nature of the Gardner phase},
such static avalanches are scale-free, and their PDF shows power-law behavior.
Note that their theoretical prediction provides the quantitative value
of the avalanche exponent $\tau$ and, importantly, argues that the
values of $\tau$ are different between systems exactly at the jamming
point ($\tau\approx 1.413$) and above jamming ($\tau=1.0$).
}

\section{Total squared displacement in the elastic regime}\label{ap:TSD}
\begin{figure}
  \includegraphics[width=\linewidth]{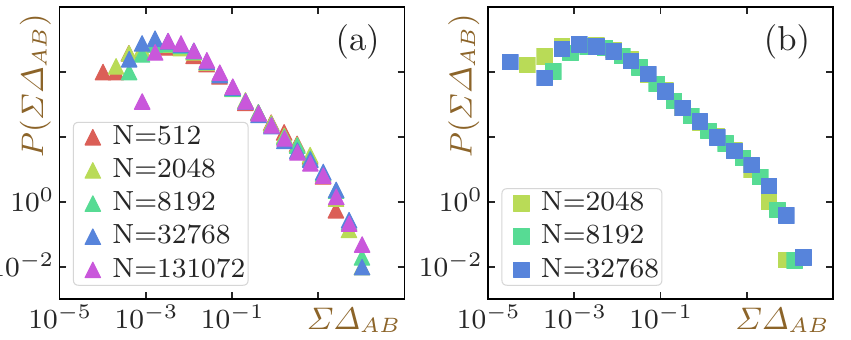}
\caption{
PDFs of the TSD $\Sigma\Delta_{AB}$.
(a) Results of the first event ensemble.
(b) Results in the elastic regime.
The different colors represent different system sizes, as shown in the
legend.
    \label{fig:TSD}}
\end{figure}

In Secs.~\ref{sec:first} and \ref{sec:transient} in the main text,
we present the PDFs of the MSDs of the avalanches of the ensembles of the first event
and of those in the elastic regime.
In this appendix, we present the PDFs of the TSD, $\Sigma\Delta_{AB}$, defined as
\begin{align}
  \Sigma\Delta_{AB}\equiv N\Delta_{AB}.
\end{align}
This gives the geometrical sizes of the events themselves.
The results are shown in Fig.~\ref{fig:TSD}.
In accordance with the avalanche sizes, the
TSDs do not show any system size dependence in the case of the first
event ensemble (Fig.~\ref{fig:TSD}(a)).
However, in the case of the elastic regime, we observe a slight system
size dependence (Fig.~\ref{fig:TSD}(b)).

%------------------------------------------------------------------------------------------
%%% Figure: TSD in transient regime
%------------------------------------------------------------------------------------------
%------------------------------------------------------------------------------------------

%
%\bibliography{Avalanche_amorphous}

%
%
\end{document}